# A Note on Location Parameter Estimation using the Weighted Hodges–Lehmann Estimator

Xuehong Gao[1], Zhijin Chen[1], Bosung Kim[2], and Chanseok Park[3]

[1]Research Institute of Macro-Safety Science, University of Science and Technology Beijing, Beijing, China
[2]School of Management, Kyung Hee University, Seoul, Republic of Korea
[3]Department of Industrial Engineering, Pusan National University, Busan, Republic of Korea

**Abstract**: Robust design is one of the main tools employed by engineers for the facilitation of the design of high-quality processes. However, most real-world processes invariably contend with external uncontrollable factors, often denoted as outliers or contaminated data, which exert a substantial distorting effect upon the computed sample mean. In pursuit of mitigating the inherent bias entailed by outliers within the dataset, the concept of weight adjustment emerges as a prudent recourse, to make the sample more representative of the statistical population. In this sense, the intricate challenge lies in the judicious application of these diverse weights toward the estimation of an alternative to the robust location estimator. Different from the previous studies, this study proposes two categories of new weighted Hodges-Lehmann (WHL) estimators that incorporate weight factors in the location parameter estimation. To evaluate their robust performances in estimating the location parameter, this study constructs a set of comprehensive simulations to compare various location estimators including mean, weighted mean, weighted median, Hodges-Lehmann estimator, and the proposed WHL estimators. The findings unequivocally manifest that the proposed WHL estimators clearly outperform the traditional methods in terms of their breakdown points, biases, and relative efficiencies.

**Keywords:** Contaminated data; Weighted Hodges-Lehmann estimator; Robustness; Location estimator.

## 1 Introduction

In statistics, the location estimation of a distribution is usually based on relatively complete and real sample data. However, in many practical cases, most processes are affected by contaminated data, which are caused by external uncontrollable factors, such as measurement errors, volatile operation conditions, etc (Park, Kim, and Wang 2022). These contaminated data within the sample have a significant influence on biasedly estimating the performance of the whole system. To enhance the reliability and performance of products, processes, and systems, robust design is usually employed in various engineering and quality assurance disciplines. In this sense, robust estimation has emerged as a critical area of research in robust design to provide more reliable and stable location parameter estimates in the face of these challenges (Gao et al. 2022; Park, Gao, and Wang 2023). In robust design, the basic assumption is that the sample comes from a normal or some other distribution. After that, the sample median, weighted median (Gao et al. 2022), and Hodges-Lehmann (HL) (Hodges and Lehmann 1963) estimators are usually considered as the alternative to the location estimator (i.e., sample mean) because they have a large breakdown point and perform well in either the presence or absence of outliers.

In the past few decades, various approaches have been proposed to deal with the robust estimation of the location parameter with the HL estimator when the sample has contamination data or outliers. For instance, Alloway Jr and Raghavachari (1991) constructed a control chart based on the HL estimator. Rousseeuw and Verboven (2002) studied several well-known robust estimators (i.e., the HL estimator and M-estimators) in very small samples. Schoonhoven et al. (2011) studied several robust location estimators for constructing the location control chart. They also analyzed the HL estimator based on the pseudo-median, which was proved to be unbiased. Park Chanseok (2016) proposed a dual quadratic response surface model to compare the joint use of various estimators, where the sample median and HL estimators were used at each design point.

As illustrated above, most studies developed their alternative robust location estimators based on the sample median and HL estimator. However, to reduce the bias of experimental data, weighting adjustment is commonly considered as a sensible remedy to make the sample more representative of the statistical population. The different weights make the alternative of the robust location estimator challenging because each of the observations cannot be treated equally. In such a case, Gao et al. (2022) combined the weights and developed a robust estimator based on the weighted median to substitute the weighted mean so that the optimal solution is not sensitive to the contaminated data or outliers within the demand for medical staff. However, when considering observations with associated weights, the conventional HL estimator proves inadequate in incorporating these weights, potentially resulting in significant deviations in the evaluation outcomes. Therefore, there is a pressing need to create an alternative version based on the conventional HL estimator, which can seamlessly integrate weights and is an aspect overlooked in prior research. As a consequence, this study aims to develop some new robust estimators given different weights in the experimental data, called weighted Hodges-Lehmann (WHL) estimators in this study, and evaluate their robust performance.

The remainder of this paper is organized as follows. In Section 2, this study presents the proposed WHL estimators with definitions. The breakdown points of the newly proposed WHL estimators are investigated in Section

3. In Section 4, a comprehensive simulation study is carried out, where the proposed location estimators are compared with previous conventional location estimators to illustrate their performances in terms of bias and relative efficiency. Finally, Section 5 concludes this study, outlining its contributions and possible directions for future work.

# 2 Methodology

It is well known that the HL estimator is a robust and nonparametric location estimator (Hodges and Lehmann 1963), which is defined as the median pairwise averages of the sample observations. Given a set of observations $x_1, x_2, \dots, x_n$, the basic Hodges-Lehmann estimator, denoted by $HL$, is given by

$$HL = \text{Median}\left(\frac{H_{ij}}{2}\right) = \text{Median}\left(\frac{x_i + x_j}{2}\right) \tag{1}$$

where both $i$ and $j$ are the index of the observations and the set $H_{ij}$ for all $i$ and $j$ is given by

$$H_{ij} = \begin{bmatrix} h_{11} & h_{12} & \cdots & h_{1n} \\ h_{21} & h_{22} & & h_{2n} \\ \vdots & & \ddots & \vdots \\ h_{n1} & h_{n2} & \cdots & h_{nn} \end{bmatrix} = \begin{bmatrix} x_1 + x_1 & x_1 + x_2 & & x_1 + x_n \\ x_2 + x_1 & x_2 + x_2 & \cdots & x_2 + x_n \\ \vdots & & \ddots & \vdots \\ x_n + x_1 & x_n + x_2 & \cdots & x_n + x_n \end{bmatrix} \tag{2}$$

As presented above, the weights were not considered in the traditional HL estimator. Therefore, new robust estimators that incorporate the weights of observations must be incorporated. Accordingly, this study proposes two new categories of WHL estimators to substitute the HL estimator in dealing with observations with weights. Particularly, the first category of WHL estimator is the median of all pairwise weighted averages of the sample observations, and the second category of WHL estimator is the weighted median of all pairwise weighted averages of the sample observations.

## 2.1 The first category of WHL estimators

As presented in Formula (1), the $\text{Median}\left(\frac{x_i+x_j}{2}\right)$ is considered a "pseudo-median" and closely related to the population median. However, after the weighting adjustment is applied to the sample observations, the weighted average needs to be used. Then we have the following definition:

**Definition 1**: Given a set of observations $x_1, x_2, \dots, x_n$ with corresponding positive weights $w_1, w_2, \dots, w_n$, such that $\sum_{i=1}^{n} w_i = 1$. The first category of WHL estimator is defined as the median of all pairwise weighted averages of the sample observations, denoted by $WHL1$, which is given by:

$$WHL1 = Median\left(\frac{L_{ij}}{w_i + w_j}\middle| w_i + w_j\right) \tag{3}$$

where

$$L_{ij} = \begin{bmatrix} \frac{w_1x_1 + w_1x_1}{w_2x_2 + w_1x_1} & \frac{w_1x_1 + w_2x_2}{w_2x_2 + w_2x_2} & \cdots & \frac{w_1x_1 + w_{n-1}x_{n-1}}{w_2x_2 + w_{n-1}x_{n-1}} & \frac{w_1x_1 + w_nx_n}{w_2x_2 + w_nx_n} \\ \vdots & & \ddots & \vdots & \\ \frac{w_{n-1}x_{n-1} + w_1x_1}{w_nx_n + w_1x_1} & \frac{w_{n-1}x_{n-1} + w_2x_2}{w_nx_n + w_2x_2} & \cdots & \frac{w_{n-1}x_{n-1} + w_{n-1}x_{n-1}}{w_nx_n + w_{n-1}x_{n-1}} & \frac{w_{n-1}x_{n-1} + w_nx_n}{w_nx_n + w_nx_n} \end{bmatrix} \tag{4}$$

Similar to the $HL$ estimator presented in (Park, Kim, and Wang 2022), the $WHL1$ can also be calculated for three cases: namely (i) $i < j$, (ii) $i \leq j$, and (iii) $\forall(i,j)$. These three versions are denoted as follows:

$$WHL1(i < j) = Median\left(\frac{w_ix_i + w_jx_j}{w_i + w_j}\middle| w_i + w_j\right), \quad \text{for } i < j \tag{5}$$

$$WHL1(i \leq j) = Median\left(\frac{w_ix_i + w_jx_j}{w_i + w_j}\middle| w_i + w_j\right), \quad \text{for } i \leq j \tag{6}$$

$$WHL1(\forall(i,j)) = Median\left(\frac{w_ix_i + w_jx_j}{w_i + w_j}\middle| w_i + w_j\right), \quad \text{for } \forall(i,j) \tag{7}$$

### 2.2 The second category of WHL estimators

Then, we derive the second category of WHL estimators that are measured by the weighted median of all pairwise weighted averages of the sample observations, and its definition is given by:

**Definition 2**: Given a set of observations $x_1, x_2, \dots, x_n$ with corresponding positive weights $w_1, w_2, \dots, w_n$, such that $\sum_{i=1}^{n} w_i = 1$. The second type of WHL estimator is defined as the weighted median of all pairwise, denoted by $WHL2$ and given by:

$$WHL2 = \text{Weighted median}\left(\frac{L_{ij}}{w_i + w_j}\middle| w_i + w_j\right) \tag{8}$$

Similar to the $HL$ estimator presented in (Park, Kim, and Wang 2022), $WHL2$ can also be calculated for three cases: (i) $i < j$, (ii) $i \leq j$, and (iii) $\forall(i,j)$. These three versions are presented as follows:

$$WHL2(i < j) = \text{Weighted median}\left(\frac{w_ix_i + w_jx_j}{w_i + w_j}\middle| w_i + w_j\right), \quad \text{for } i < j \tag{9}$$

$$WHL2(i \leq j) = \text{Weighted median}\left(\frac{w_i x_i + w_j x_j}{w_i + w_j}\middle| w_i + w_j\right), \quad \text{for } i \leq j \tag{10}$$

$$WHL2[\forall(i,j)] = \text{Weighted median}\left(\frac{w_i x_i + w_j x_j}{w_i + w_j}\middle| w_i + w_j\right), \quad \text{for } \forall(i,j) \tag{11}$$

where the Weighted median () is the function to calculate the 50% weighted percentile.

# 3 Breakdown point

In the context of robust design, the robustness of an estimator is usually evaluated using the well-known breakdown point criterion (Donoho and Huber 1983; Park, Kim, and Wang 2022), which quantifies the maximum proportion of outliers that the estimator can endure before it breaks down. For example, the breakdown points of the sample mean and sample median for the location parameter are 0 and 0.5, respectively. Since the breakdown point can generally be written as a function of the sample size (e.g., usually denoted by $n$), this study also develops the finite-sample breakdown-point function for the various location estimators mentioned above. In addition to the breakdown point criterion, this study also utilizes bias and relative efficiency to evaluate the performance of the newly proposed WHL estimators.

### 3.1 Breakdown point of $WHL1$

After comparing the $WHL1$ estimators with the HL estimator, it was found that the only difference between them is that the $WHL1$ estimators apply a weighted average in each pair. In this sense, the $WHL1$ estimators have the same breakdown point as the HL estimator for the three corresponding cases. Based on our previous work (Ouyang Linhan 2019; Park, Kim, and Wang 2022), we summarize the breakdown points for the three cases presented in Formulars (4)-(6) given a sample size $n$. Thus, we have the corresponding breakdown points for the above three $WHL1$ cases, which are provided as follows.

**Theorem 1:** The breakdown points for the above three $WHL1$ cases are:

$$BP_{WHL1(i<j)} = \frac{\left\lfloor n - \frac{1}{2} - \sqrt{\left(n - \frac{1}{2}\right)^2 - 2\left\lfloor \frac{n^2 - n - 2}{4} \right\rfloor} \right\rfloor}{n} \tag{12}$$

| | |
|---|---|
| $$BP_{WHL1(i\leq j)}=\frac{\left\lfloor n+\frac{1}{2}-\sqrt{\left(n+\frac{1}{2}\right)^2-2\left\lfloor\frac{n^2+n-2}{4}\right\rfloor}\right\rfloor}{n}$$ | (13) |
| $$BP_{WHL1\forall(i,j))}=\frac{\left\lfloor n-\sqrt{n^2-\left\lfloor\frac{n^2-1}{2}\right\rfloor}\right\rfloor}{n}$$ | (14) |

**Proof**: To validate the breakdown points for the above three $WHL1$ cases, it is required to account for the proportion of incorrect or arbitrarily large pairwise caused by the outliers. With **Definition 1**, given a contaminated observation $x_i{}'$ or $x_j{}'$, we have

| | |
|---|---|
| $$WHL1'=\text{Median}\left(\frac{w_i x_i{}'+w_j x_j}{w_i+w_j}\middle\vert w_i+w_j\right)\text{ or }WHL1'=\text{Median}\left(\frac{w_i x_i+w_j x_j{}'}{w_i+w_j}\middle\vert w_i+w_j\right)$$ | (15) |

According to the above equation, it is easily seen that the pairwise is directly affected by its observation values rather than the weights. In addition, the traditional $HL$ estimator is a special case of the $WHL1$ estimator when the weights are equal. Since the pairwise in the traditional $HL$ estimator is only influenced by its observation values, the $WHL1$ estimator has the same breakdown points as the traditional $HL$ estimator provided by Park, Kim, and Wang (2022). Then we complete the proof for validating the breakdown points for the $WHL1$ estimator.

To present the robustness property of the $WHL1$ estimator, several robust estimators are compared given different sample sizes (i.e., from 1 to 50), and a plot of these values is provided in Fig. 1. Particularly, the breakdown point for the $WHL1$ is provided first, which is the same as that of the $HL$ estimator estimated by Park, Kim, and Wang (2022). The breakdown point of the median is also shown in Fig. 1. In terms of the weighted median (see the blue and green lines in Fig. 1), the lower and upper bounds of the breakdown point are presented as they are strongly correlated with the observation weights. The $WHL1$'s breakdown point is related to the number of pairwise rather than observations, making the breakdown point of the $WHL1$ estimator different from that of the traditional weighted median.

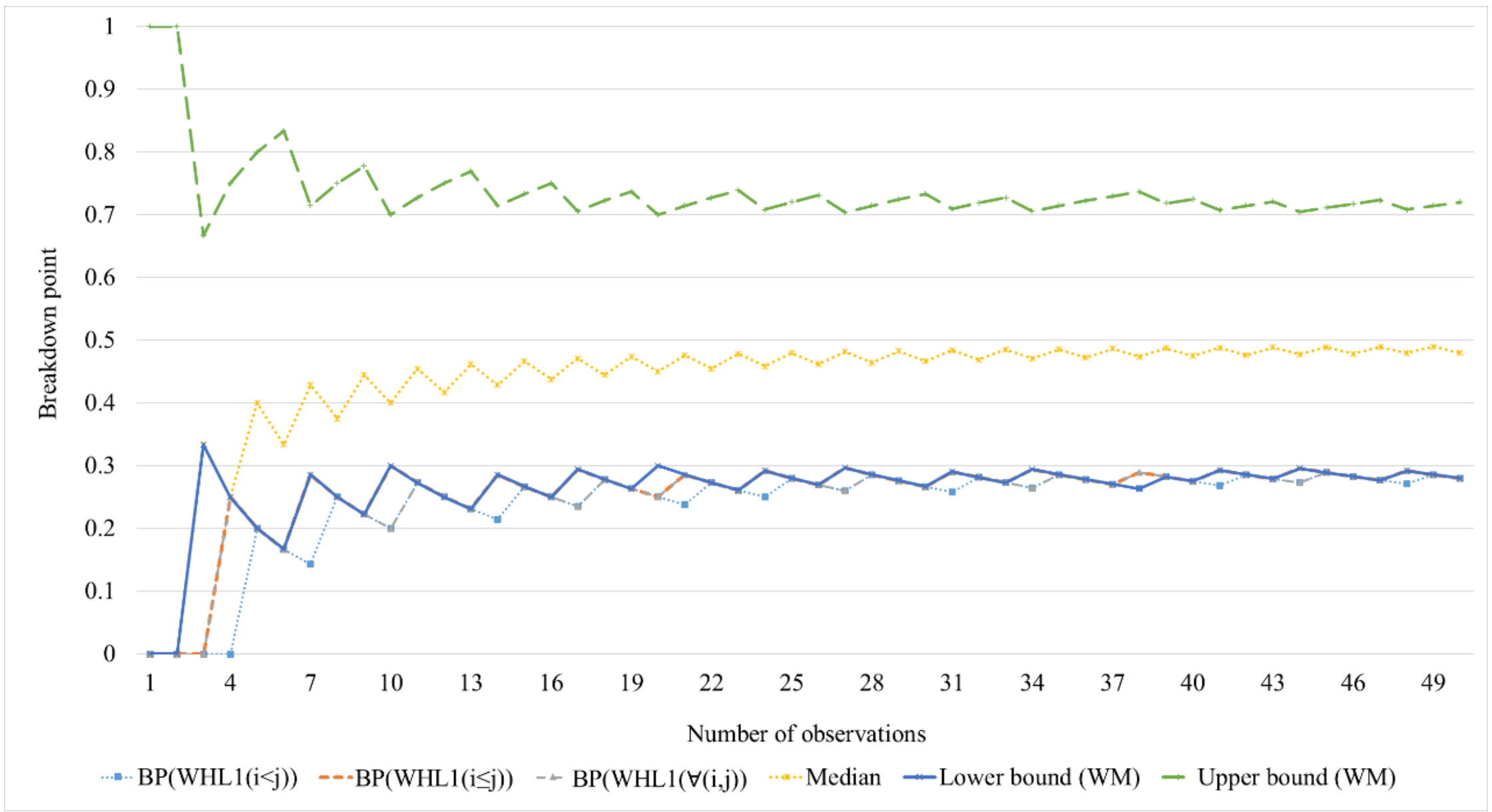

Fig. 1. Illustration of breakdown points of robust estimators

### 3.2 Breakdown point of $WHL2$

Furthermore, this study examines the breakdown point of the $WHL2$ estimator. Unlike $WHL1$, the $WHL2$ estimator considers the weighted median for all pairwise rather than the sample size. In this sense, the breakdown point of the $WHL2$ estimator is highly correlated with the outlier weights. It is well known that the weighted median is the 50% weighted percentile (Gao et al. 2022). Because the summation for all pairwise sample weights is greater than 1, it is necessary to renormalize the weights for the above three versions. To represent the pairwise weight of sample observations $i$ and $j$, three cases are considered and given as follows:

| | |
|---|---|
| $W_{ij}(i < j) = (w_i + w_j)/\sum_{i}^{n}\sum_{j}^{i-1}(w_i + w_j), \quad for\ i < j$ | (16) |
| $W_{ij}(i \le j) = (w_i + w_j)/\sum_{i}^{n}\sum_{j}^{i}(w_i + w_j), \quad for\ i \le j$ | (17) |
| $W_{ij}[\forall(i,j)] = (w_i + w_j)/\sum_{i}^{n}\sum_{j}^{n}(w_i + w_j), \quad for\ \forall(i,j)$ | (18) |

According to the equations (16)-(18), the total numbers of elements/pairwise for the aforementioned three cases are (i) $m = (n^2 - n)/2$, (ii) $m = (n^2 + n)/2$, and (iii) $m = n^2$. In the following, we derive the breakdown point for each case.

**(i) Case 1**

After applying the weighting adjustment to the sample, the 50% weighted percentile is calculated. For convenience, let $z_1, z_2, \ldots, z_m$ represent the ascending order statistics with corresponding weights $\omega_{(1)}, \omega_{(2)}, \ldots, \omega_{(m)}$ such that $\sum_{i=1}^{m} \omega_{(i)} = 1$, where $m = (n^2 - n)/2$ is the total number of elements. Thus, the weighted median in this case, denoted by $WM1$, can be obtained as follows:

$$WM1 = \inf\left\{z: \sum_{i=1}^{k} \omega_{(i)} > \frac{1}{2}\right\} = \sup\left\{z: \sum_{i=1}^{k} \omega_{(i)} \leq \frac{1}{2}\right\} = z_{(k+1)} \quad (19)$$

For more details on how to obtain the weighted median, readers are referred to previous studies (Gao 2020; Gao et al. 2022).

**Theorem 2**: Given a set of observations $x_1, x_2, \ldots, x_n$ with corresponding positive weights $w_1, w_2, \ldots, w_n$, $WHL2(i < j)$ can be obtained using equation (9). After a combination of reordering and reweighting operations, ascending order statistics $z_1, z_2, \ldots, z_m$ with corresponding weights $\omega_{(1)}, \omega_{(2)}, \ldots, \omega_{(m)}$ such that $\sum_{i=1}^{m} \omega_{(i)} = 1$ can be obtained. Thus, the breakdown point for $WHL2(i < j)$ is given as follows:

$$BP_{WHL2(i<j)} = \frac{\max\left\{k \leq \frac{n^2 - n}{2} - 1 : \sum_{i=1}^{k} \omega_{(i)} < \frac{1}{2}\right\}}{\frac{n^2 - n}{2}} \quad (20)$$

**Proof:**

From **Definition 2**, it was found that $WHL2(i < j)$ is the weighted median of all pairwise rather than the traditional median. In other words, the robustness of $WHL2(i < j)$ is related to the pairwise weights. To derive the breakdown point for $WHL2(i < j)$, it is required to know the number of pairwise while i < j, which is given by

$$m = \frac{n^2 - n}{2} \quad (21)$$

Combined with Definition in (27) proposed by Gao et al. (2022), the breakdown point for $WHL2(i < j)$ is given by

$$BP_{WHL2(i<j)} = \frac{\max\left\{k \leq m - 1 : \sum_{i=1}^{k} \omega_{(i)} < \frac{1}{2}\right\}}{m} = \frac{\max\left\{k \leq \frac{n^2 - n}{2} - 1 : \sum_{i=1}^{k} \omega_{(i)} < \frac{1}{2}\right\}}{\frac{n^2 - n}{2}} \quad (22)$$

Thus, we complete the proof for Theorem 2.

**(ii) Case 2**

Similarly, let $z_1, z_2, \ldots, z_m$ represent the ascending order statistics with corresponding weights $\omega_{(1)}, \omega_{(2)}, \ldots, \omega_{(m)}$ such that $\sum_{i=1}^{m} \omega_{(i)} = 1$, where $m = \frac{n^2+n}{2}$ is the total number of elements/ pairwise in Case 2. Then, the weighted median, in this case, denoted by $WM2$, can be obtained as follows:

$$WM2 = \inf\left\{z: \sum_{i=1}^{k} \omega_{(i)} > \frac{1}{2}\right\} = \sup\left\{z: \sum_{i=1}^{k} \omega_{(i)} \leq \frac{1}{2}\right\} = z_{(k+1)}, \qquad k \leq \frac{n^2+n}{2} \tag{23}$$

**Theorem 3**. Given a set of observations $x_1, x_2, \ldots, x_n$ with corresponding positive weights $w_1, w_2, \ldots, w_n$, $WHL2(i \leq j)$ can be obtained using equation (10). After a combination of reordering and reweighting operations, ascending order statistics $z_1, z_2, \ldots, z_m$ with corresponding weights $\omega_{(1)}, \omega_{(2)}, \ldots, \omega_{(m)}$ such that $\sum_{i=1}^{m} \omega_{(i)} = 1$ can be obtained. Thus, the breakdown point for $WHL2(i \leq j)$ is given as follows:

$$BP_{WHL2(i\leq j)} = \frac{\max\left\{k \leq \frac{n^2+n}{2} - 1: \sum_{i=1}^{k} \omega_{(i)} < \frac{1}{2}\right\}}{\frac{n^2+n}{2}} \tag{24}$$

**Proof:** From Definition 2 presented in (10), $WHL2(i \leq j)$ is also the weighted median of all pairwise rather than the traditional median. To derive the breakdown point for $WHL2(i \leq j)$, it is required to know the number of pairwise while $i \leq j$, which is given by

$$m = \frac{n^2+n}{2} \tag{25}$$

Combined with Definition in (27) proposed by Gao et al. (2022), the breakdown point for $WHL2(i \leq j)$ is given by

$$BP_{WHL2(i\leq j)} = \frac{\max\left\{k \leq m - 1: \sum_{i=1}^{k} \omega_{(i)} < \frac{1}{2}\right\}}{m} = \frac{\max\left\{k \leq \frac{n^2+n}{2} - 1: \sum_{i=1}^{k} \omega_{(i)} < \frac{1}{2}\right\}}{\frac{n^2+n}{2}} \tag{26}$$

Thus, we complete the proof for Theorem 3.

**(iii) Case 3**

Similarly, let $z_1, z_2, \ldots, z_m$ represent the ascending order statistics with corresponding weights $\omega_{(1)}, \omega_{(2)}, \ldots, \omega_{(m)}$ such that $\sum_{i=1}^{m} \omega_{(i)} = 1$, where $m = n^2$ is the total number of elements/pairwise in Case 3. Then, the weighted median, in this case, denoted by $WM3$, can be obtained as follows:

$$WM3 = \inf\left\{z: \sum_{i=1}^{k} \omega_{(i)} > \frac{1}{2}\right\} = \sup\left\{z: \sum_{i=1}^{k} \omega_{(i)} \leq \frac{1}{2}\right\} = z_{(k+1)}, \qquad k \leq n^2 - 1 \tag{27}$$

**Theorem 4**. Given a set of observations $x_1, x_2, \dots, x_n$ with corresponding positive weights $w_1, w_2, \dots, w_n$, $WHL2[\forall(i,j)]$ can be obtained using Equation (11). After a combination of reordering and reweighting operations, ascending order statistics $z_1, z_2, \dots, z_m$ with corresponding weights $\omega_{(1)}, \omega_{(2)}, \dots, \omega_{(m)}$ such that $\sum_{i=1}^{m} \omega_{(i)} = 1$ can be obtained. Thus, the breakdown point for $WHL2[\forall(i,j)]$ is given as follows:

$$BP_{WHL2[\forall(i,j)]} = \frac{\max\left\{k \leq n^2 - 1: \sum_{i=1}^{k} \omega_{(i)} < \frac{1}{2}\right\}}{n^2} \tag{28}$$

From Definition 2 presented in (11), $WHL2[\forall(i,j)]$ is also the weighted median of all pairwise rather than the traditional median. To derive the breakdown point for $WHL2[\forall(i,j)]$, it is required to know the number of pairwise while $\forall(\mathrm{i},\mathrm{j})$, which is given by

$$m = n^2 \tag{29}$$

Combined with the Definition in (27) proposed by Gao et al. (2022), the breakdown point for $WHL2[\forall(i,j)]$ is given by

$$BP_{WHL2[\forall(i,j)]} = \frac{\max\left\{k \leq m - 1: \sum_{i=1}^{k} \omega_{(i)} < \frac{1}{2}\right\}}{m} = \frac{\max\left\{k \leq n^2 - 1: \sum_{i=1}^{k} \omega_{(i)} < \frac{1}{2}\right\}}{n^2} \tag{30}$$

Thus, we complete the proof for Theorem 4.

### 3.3 Breakdown point comparison

Because the breakdown point of the $WHL2$ estimators is strongly correlated with the weights, it is difficult to derive the specific formulation of the breakdown point. Thus, this study considered two cases (best and worst) to obtain the upper- and lower-bound breakdown points.

#### (1) Best case

Suppose that the set of $z_1, z_2, \dots, z_m$ is the order statistic of $\omega^u{}_{(1)}, \omega^u{}_{(2)}, \dots, \omega^u{}_{(m)}$ such that $\omega^u{}_{(1)} \leq \omega^u{}_{(2)} \leq, \dots, \leq \omega^u{}_{(m)}$ and $\sum_{i=1}^{m} \omega^u{}_{(i)} = 1$. In the best case, the set of contaminated pairwise has smaller weights. Thus, the upper-bound breakdown point is given by

$$BP^u = \frac{\max\left\{k \le m-1: \sum_{i=1}^{k} \omega^u{}_{(i)} < \frac{1}{2}\right\}}{m} \quad (31)$$

Note that with the definition $f(0) = 0$ and $f(x) = \sum_{i=1}^{x} \omega^u{}_{(i)}$ for $x = 1, 2, \dots, m, f(x)$ is a discrete convex function. For more details, please refer to **Lemma 1** of Gao et al. (2022).

**(2) Worst case**

We then consider the worst case, in which the lower bound of the breakdown point for $WHL2$ can be obtained. Suppose that the set of $z_1, z_2, \dots, z_m$ is the order statistics of $\omega^l{}_{(1)}, \omega^l{}_{(2)}, \dots, \omega^l{}_{(m)}$ such that $\omega^l{}_{(1)} \ge \omega^l{}_{(2)} \ge, \dots, \ge \omega^l{}_{(m)}$ and $\sum_{i=1}^{m} \omega^l{}_{(i)} = 1$. When the set of contaminated pairwise has larger weights, the lower-bound breakdown point can be obtained as follows:

$$BP^l = \frac{\max\left\{k \le m-1: \sum_{i=1}^{k} \omega^l{}_{(i)} < \frac{1}{2}\right\}}{m} \quad (32)$$

Note that with definition $f(0) = 0$ and $f(x) = \sum_{i=1}^{x} \omega^l{}_{(i)}$ for $x = 1, 2, \dots, m, f(x)$ is a discrete concave function. For more details, please refer to **Lemma 2** of Gao et al. (2022).

Thus, the breakdown point of the $WHL2$ estimator can be roughly estimated. Particularly, to estimate the lower-bound breakdown point, the worst case is considered with the assumption $\omega^l{}_{(i+1)} - \omega^l{}_{(i)} = \omega^l{}_{(i)} - \omega^l{}_{(i-1)} > 0$. To estimate the upper-bound breakdown point, the best case is considered under the assumption $\omega^l{}_{(i-1)} - \omega^l{}_{(i)} = \omega^l{}_{(i)} - \omega^l{}_{(i+1)} > 0$. To illustrate the robustness of the newly proposed $WHL2$ estimator, different methods are compared, and Table 1 provides the breakdown points when the sample size increases from 1 to 20. As shown in Table 1, the breakdown point of the proposed $WHL1$ estimator is the same as that of the traditional $HL$ estimator. However, the breakdown point of the $WHL2$ estimator is different from that of the $WHL1$ estimator owing to the intrinsic weights.

Table 1 Properties (i.e., the number of pairwise and breakdown points) of the newly proposed WHL estimators

| **Sample size** | **BP (Median)** | **BP (Weighted Median)** | **Number of pariwise+BP ($WHL1$)** | | | **Number of pariwise+BP ($WHL2$)** | | |
|---|---|---|---|---|---|---|---|---|
| | | | $i<j$ | $i\le j$ | $\forall(i,j)$ | $i<j$ | $i\le j$ | $\forall(i,j)$ |
| 1 | 0 | (0.000,1.000) | 0+0.000 | 1+0.000 | 1+0.000 | 0+(0.000,1.000) | 1+(0.000,1.000) | 1+(0.000,1.000) |
| 2 | 0 | (0.000,1.000) | 1+0.000 | 3+0.000 | 4+0.000 | 1+(0.000,1.000) | 3+(0.333,0.667) | 4+(0.250,0.750) |
| 3 | 1/3 | (0.333,0.667) | 3+0.000 | 6+0.000 | 9+0.000 | 3+(0.333,0.667) | 6+(0.167,0.833) | 9+(0.222,0.778) |
| 4 | 1/4 | (0.250,0.750) | 6+0.000 | 10+0.250 | 16+0.250 | 6+(0.167,0.833) | 10+(0.300,0.700) | 16+(0.250,0.750) |
| 5 | 0.4 | (0.200,0.800) | 10+0.200 | 15+0.200 | 25+0.200 | 10+(0.300,0.700) | 15+(0.267,0.733) | 25+(0.280,0.720) |
| 6 | 0.333 | (0.167,0.833) | 15+0.167 | 21+0.167 | 36+0.167 | 15+(0.267,0.733) | 21+(0.286,0.714) | 36+(0.278,0.722) |
| 7 | 0.429 | (0.286,0.714) | 21+0.143 | 28+0.286 | 49+0.286 | 21+(0.286,0.714) | 28+(0.286,0.714) | 49+(0.286,0.714) |
| 8 | 0.375 | (0.250,0.750) | 28+0.250 | 36+0.250 | 64+0.250 | 28+(0.286,0.714) | 36+(0.278,0.722) | 64+(0.281,0.719) |
| 9 | 0.444 | (0.222,0.778) | 36+0.222 | 45+0.222 | 81+0.222 | 36+(0.278,0.722) | 45+(0.289,0.711) | 81+(0.272,0.728) |
| 10 | 0.4 | (0.300,0.700) | 45+0.200 | 55+0.300 | 100+0.200 | 45+(0.289,0.711) | 55+(0.291,0.709) | 100+(0.300,0.700) |
| 11 | 0.455 | (0.273,0.727) | 55+0.273 | 66+0.273 | 121+0.273 | 55+(0.291,0.709) | 66+(0.288,0.712) | 121+(0.289,0.711) |
| 12 | 0.417 | (0.250,0.750) | 66+0.250 | 78+0.250 | 144+0.250 | 66+(0.288,0.712) | 78+(0.282,0.718) | 144+(0.292,0.708) |
| 13 | 0.462 | (0.231,0.769) | 78+0.231 | 91+0.231 | 169+0.231 | 78+(0.282,0.718) | 91+(0.286,0.714) | 169+(0.290,0.710) |
| 14 | 0.429 | (0.286,0.714) | 91+0.214 | 105+0.286 | 196+0.286 | 91+(0.286,0.714) | 105+(0.286,0.714) | 196+(0.291,0.709) |
| 15 | 0.467 | (0.267,0.733) | 105+0.267 | 120+0.267 | 225+0.267 | 105+(0.286,0.714) | 120+(0.292,0.708) | 225+(0.293,0.707) |
| 16 | 0.438 | (0.250,0.750) | 120+0.250 | 136+0.250 | 256+0.250 | 120+(0.292,0.708) | 136+(0.287,0.713) | 256+(0.293,0.707) |
| 17 | 0.471 | (0.294,0.706) | 136+0.235 | 153+0.294 | 289+0.235 | 136+(0.287,0.713) | 153+(0.288,0.712) | 289+(0.291,0.709) |
| 18 | 0.444 | (0.278,0.722) | 153+0.278 | 171+0.278 | 324+0.278 | 153+(0.288,0.712) | 171+(0.292,0.708) | 324+(0.293,0.707) |
| 19 | 0.474 | (0.263,0.737) | 171+0.263 | 190+0.263 | 361+0.263 | 171+(0.292,0.708) | 190+(0.289,0.711) | 361+(0.291,0.709) |
| 20 | 0.450 | (0.300,0.700) | 190+0.250 | 210+0.250 | 400+0.250 | 190+(0.289,0.711) | 210+(0.290,0.710) | 400+(0.293,0.708) |

In addition, a visual illustration is also provided in Fig. 2, in which the breakdown points of the three different cases of $WHL1$ are separately compared with the traditional methods. It is easy to observe that given a specific sample size, the median, $HL$, and $WHL1$ have a fixed breakdown point. However, the breakdown points of the weighted median and $WHL2$ belong to a certain range owing to their intrinsic weights [see Fig. 2 (a)–(c)]. Interestingly, $WHL2$ has a stable range of breakdown points when the sample size increases from 1 to 20, compared with the weighted median. The lower-bound breakdown point of $WHL2$ is relatively higher than that of the other methods, except for the median.

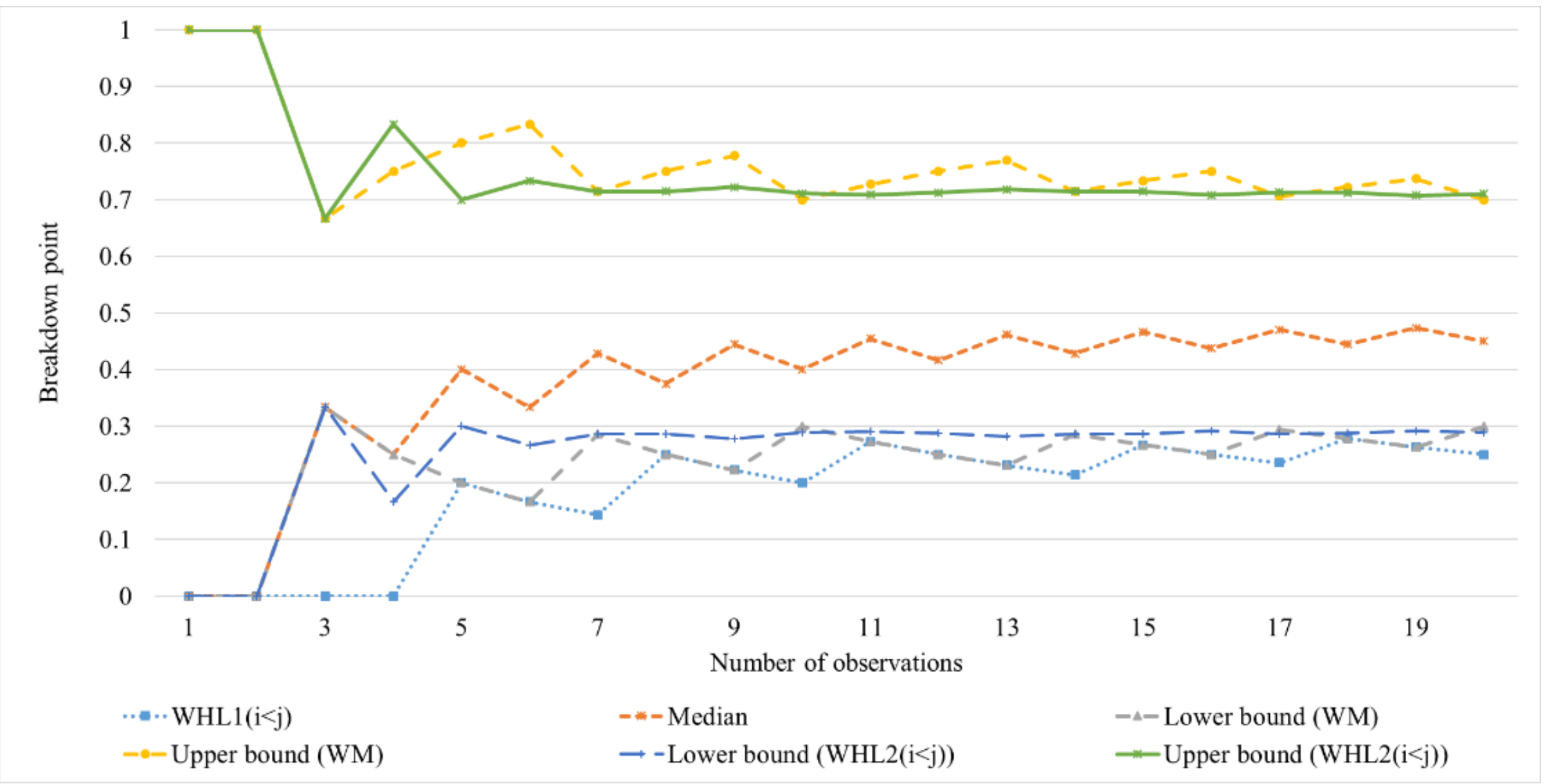

(a) $i < j$

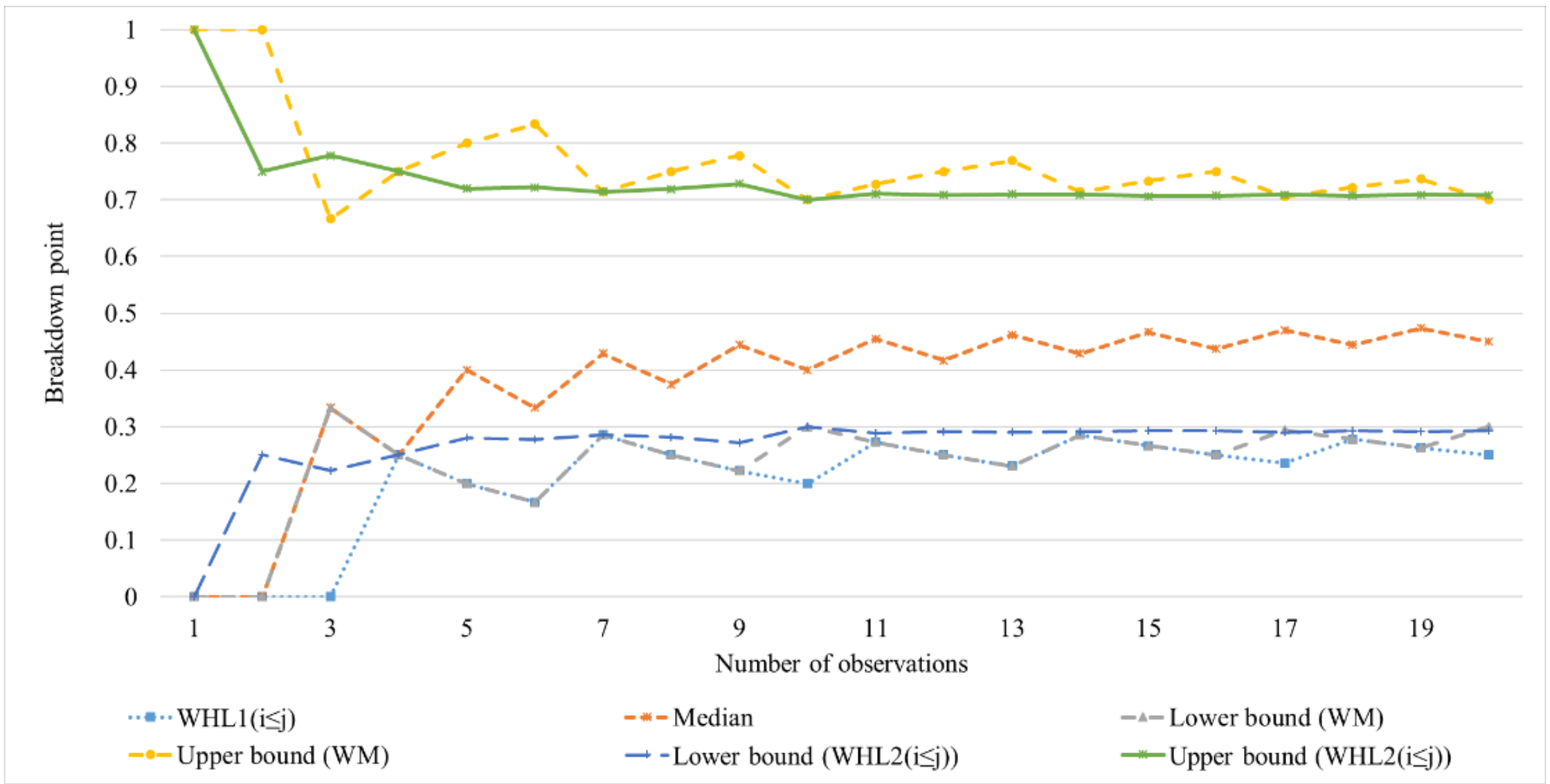

(b) $i \leq j$

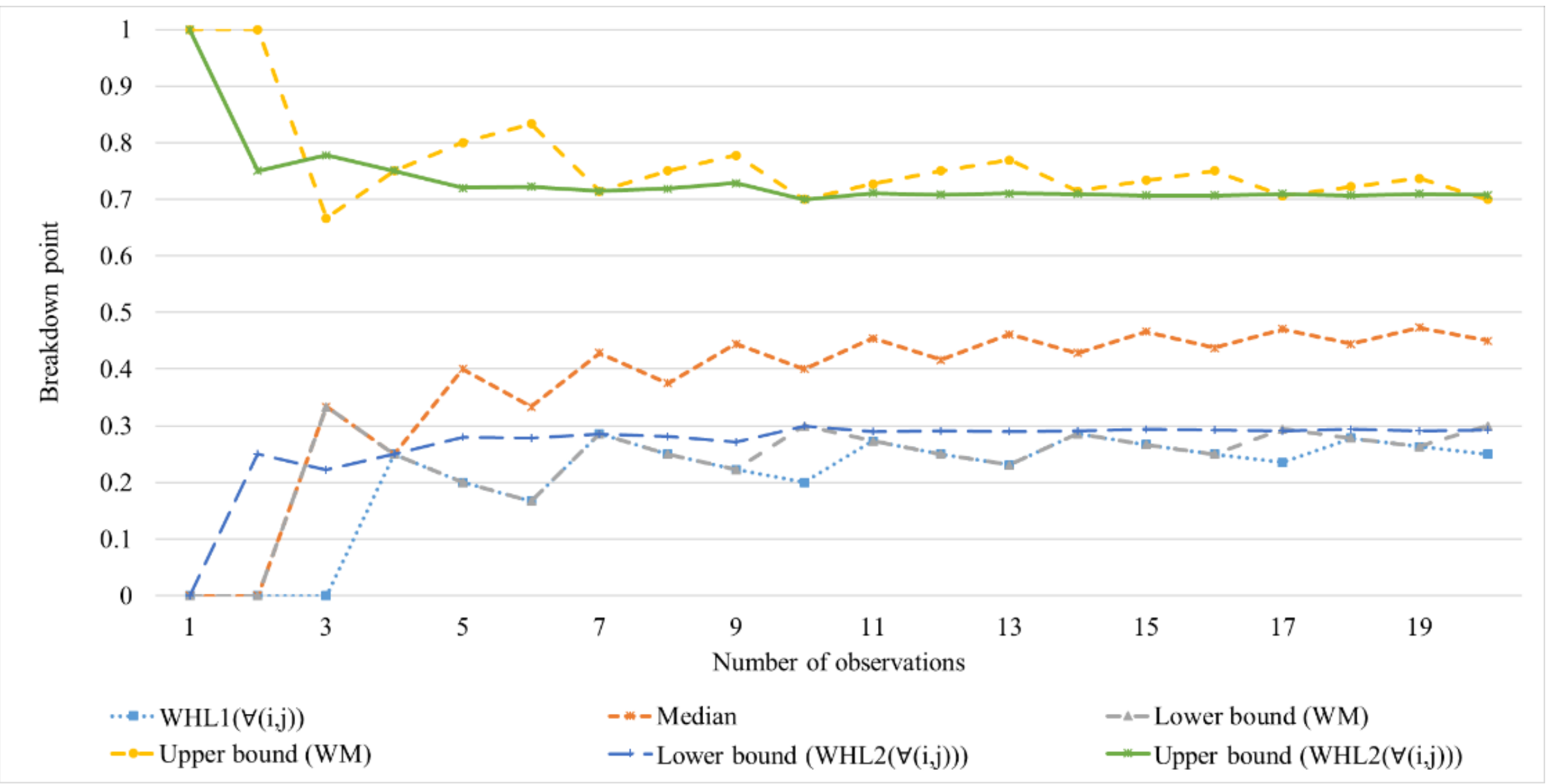

(c) $\forall(i,j)$

Fig. 2. Comparison of breakdown points under different robust estimators

# 4 Simulation study

## 4.1 Evaluation criterion

To evaluate the performances of the proposed WHL estimator, the bias and relative efficiency are evaluated and compared with the conventional location estimators. In this study, the bias is the difference between the expected value and the true value of the parameter being estimated, which is defined as

$$\text{Bias}\left(\hat{\theta},\theta\right) = \left|\hat{\theta}-\theta\right| \tag{33}$$

where $\theta$ is the weighted mean and $\hat{\theta}$ is the alternative location estimator in this study.

Efficiency (Serfling 2011) is another method to measure the quality of an estimator in the experimental design, and the relative efficiency of the two procedures is the ratio of their efficiencies. Here, the relative efficiency is used as a metric for comparing the effectiveness of the two estimators, which has been widely used in previous studies (Lehmann 2004; Park and Leeds 2016; Gao and Cui 2021; Park, Kim, and Wang 2022; Park, Gao, and Wang 2023). The relative efficiency of $\theta$ and $\hat{\theta}$ is defined as

$$\text{Relative efficiency}\left(\hat{\theta},\theta\right) = \frac{\text{Var}(\theta)}{\text{Var}\left(\hat{\theta}\right)} \times 100\% \tag{34}$$

where $\theta$ is often a reference or baseline estimator. In this study, $\text{Var}(\theta)$ is the weighted variance of the sample, which is given by

$$\mathrm{Var}(\theta) = \frac{\sum_i (w_i \sigma_i)^2}{(\sum_i w_i)^2} \quad (35)$$

### 4.2 Illustrative examples

In this study, 6 samples with different sample sizes and weighted-value distributions are first carried out. The simulations associated with computations are performed using the R language. For the sample with the sample size $n$, the observation $i$ is randomly generated from the normal distribution $N(\mu_i, \sigma_i)$ with a given weight $w_i$. To estimate the bias and relative efficiency, we replicate the sample $10{,}000$ times, which results in $10{,}000$ estimated locations for the proposed WHL estimator.

In Sample 1, the weights are the same as the normalized sample mean values. Here, the sample size is $n = 4$, and there are four mean values (i.e., 4, 3, 2, and 1) with the corresponding standard deviations (i.e., 10, 5, 10, and 5) and the corresponding weights (i.e., 4/10, 3/10, 2/10, and 1/10). The dataset has a weighted mean of $\sum_i w_i \mu_i = 2$ and a weighted variance of $\sum_i (w_i \sigma_i)^2 = 15$. The results (see Table 2) of biases and relative efficiencies are calculated and compared by using weighted mean, weighted median, and the proposed WHL estimator. We use the traditional method of the weighted mean as the baseline method by calculating the relative efficiencies of other methods. Therefore, the relative efficiency of the weighted mean is always around 100%. The weighted mean and the proposed WHL estimator have a smaller bias than the weighted median except for $WHL2(i < j)$. In terms of relative efficiency, the weighted mean and the proposed WHL estimator also outperform the weighted median.

Table 2. Results of bias and relative efficiency in different methods for Sample 1

| Measures | WM | WMD | $WHL1$ | | | $WHL2$ | | |
|---|---|---|---|---|---|---|---|---|
| | | | $i<j$ | $i \le j$ | $\forall(i,j)$ | $i<j$ | $i \le j$ | $\forall(i,j)$ |
| Variance | 15.4898 | 16.9559 | 14.8596 | 14.0680 | 14.0680 | 16.5168 | 14.3338 | 15.1130 |
| Bias | 0.004.1 | 0.3877 | 0.0276 | 0.1996 | 0.1996 | 0.4193 | 0.1936 | 0.2389 |
| Relative efficiency | 96.8473 | 87.6959 | 100.9500 | 106.3350 | 89.8690 | 89.8690 | 104.3850 | 98.8887 |

WM: weighted mean; WMD: weighted median.

In addition to Sample 1, more examples (i.e., Samples 2-6) are provided and illustrated in Table 3, where the observation $i$ is randomly generated from the normal distribution $N(\mu_i, \sigma_i)$ with a given weight $w_i$. With Samples 2-6, the biases and relative efficiencies are tested and compared when the sample size $n$ goes from 3 to 15. The simulation results of biases for samples 2-6 are presented in Tables 4-8, respectively. In addition, the results of relative efficiencies for Samples 2-6 are presented in Tables 9-13, respectively.

Table 3. Description of Samples 2-6

| Sample No. | Mean | Standard deviation | weight |
|---|---|---|---|
| 2 | $1, 2, \ldots, \mu_i, \ldots, n$ | $1, 2, \ldots, \sigma_i, \ldots, n$ | $1/n$ |
| 3 | $n, \ldots, \mu_i, \ldots, 2, 1$ | $1, 2, \ldots, \sigma_i, \ldots, n$ | $1/n$ |
| 4 | $1, 2, \ldots, \mu_i, \ldots, n$ | $1, 2, \ldots, \sigma_i, \ldots, n$ | $1, 1/2, \ldots, w_i, \ldots, 1/(n-1), 1/n$ |
| 5 | $n, \ldots, \mu_i, \ldots, 2, 1$ | $1, 2, \ldots, \sigma_i, \ldots, n$ | $1, 1/2, \ldots, w_i, \ldots, 1/(n-1), 1/n$ |
| 6 | 5 | $1, 2, \ldots, \sigma_i, \ldots, n$ | $1, 1/2^2, \ldots, w_i, \ldots, 1/(n-1)^2, 1/n^2$ |

As shown in Tables 4-8, the biases are obtained for the methods of weighted mean, weighted median, and the proposed WHL estimator. Those methods have different performances of biases. Specifically, in Table 4, the biases obtained by using the different methods are quite similar. In Tables 5-8, the proposed WHL estimators (i.e., $WHL1$ and $WHL2$) have smaller biases than the weighted median. It is also obvious that $WHL2$ estimators have smaller biases than $WHL1$ estimators and weighted median in Table 6, whereas $WHL1$ estimators have smaller biases than $WHL2$ estimators only in Table 7. Generally, the $WHL2$ estimators need to be preferred to estimate the location parameter.

Table 4. Results of biases given different sample sizes for sample 2

| Sample size $n$ | WM | WMD | $WHL1$ | | | $WHL2$ | | |
|---|---|---|---|---|---|---|---|---|
| | | | $i < j$ | $i \le j$ | $i < j$ | $i \le j$ | $i < j$ | $i \le j$ |
| 3 | 0.005746 | 0.001185 | 0.008027 | 0.004606 | 0.008027 | 0.008027 | 0.004606 | 0.008027 |
| 4 | 0.001980 | 0.002878 | 0.001980 | 0.001881 | 0.001881 | 0.001980 | 0.001881 | 0.001881 |
| 5 | 0.004424 | 0.000997 | 0.004668 | 0.004130 | 0.004130 | 0.004668 | 0.004130 | 0.004130 |
| 6 | 0.001634 | 0.001739 | 0.000963 | 0.000716 | 0.000840 | 0.000963 | 0.000716 | 0.000840 |
| 7 | 0.001537 | 0.001743 | 0.001424 | 0.002450 | 0.002449 | 0.001424 | 0.002450 | 0.002449 |
| 8 | 0.001554 | 0.000187 | 0.001961 | 0.001416 | 0.001835 | 0.001961 | 0.001416 | 0.001835 |
| 9 | 0.002879 | 0.001483 | 0.002535 | 0.002483 | 0.002674 | 0.002535 | 0.002483 | 0.002674 |
| 10 | 0.002253 | 0.000377 | 0.002292 | 0.001530 | 0.001977 | 0.002292 | 0.001530 | 0.001977 |
| 11 | 0.001037 | 0.002471 | 0.001857 | 0.001638 | 0.001650 | 0.001857 | 0.001638 | 0.001650 |
| 12 | 0.002499 | 0.003920 | 0.002611 | 0.002519 | 0.002499 | 0.002611 | 0.002519 | 0.002499 |
| 13 | 0.002763 | 0.002294 | 0.002455 | 0.002300 | 0.002497 | 0.002455 | 0.002300 | 0.002497 |
| 14 | 0.000832 | 0.002610 | 0.002234 | 0.002321 | 0.002255 | 0.002234 | 0.002321 | 0.002255 |
| 15 | 0.002910 | 0.003007 | 0.002887 | 0.002785 | 0.002862 | 0.002887 | 0.002785 | 0.002862 |

WM: weighted mean; WMD: weighted median.

Table 5. Results of biases given different sample sizes for sample 3

| Sample size $n$ | WM | WMD | $WHL1$ | | | $WHL2$ | | |
|---|---|---|---|---|---|---|---|---|
| | | | $i < j$ | $i \le j$ | $i < j$ | $i \le j$ | $i < j$ | $i \le j$ |
| 3 | 0.002546 | 0.219605 | 0.113621 | 0.052992 | 0.113621 | 0.113621 | 0.052992 | 0.113621 |
| 4 | 0.009528 | 0.269012 | 0.009528 | 0.172761 | 0.172761 | 0.009528 | 0.172761 | 0.172761 |
| 5 | 0.020621 | 0.499740 | 0.141553 | 0.185723 | 0.185723 | 0.141553 | 0.185723 | 0.185723 |
| 6 | 0.005601 | 0.594033 | 0.144118 | 0.247098 | 0.195608 | 0.144118 | 0.247098 | 0.195608 |
| 7 | 0.022999 | 0.771115 | 0.192953 | 0.380594 | 0.312048 | 0.192953 | 0.380594 | 0.312048 |
| 8 | 0.000723 | 0.887979 | 0.323730 | 0.416543 | 0.376729 | 0.323730 | 0.416543 | 0.376729 |
| 9 | 0.008385 | 1.088586 | 0.359626 | 0.473124 | 0.420799 | 0.359626 | 0.473124 | 0.420799 |
| 10 | 0.022444 | 1.178590 | 0.389864 | 0.530114 | 0.452349 | 0.389864 | 0.530114 | 0.452349 |
| 11 | 0.007490 | 1.358652 | 0.476312 | 0.593880 | 0.539576 | 0.476312 | 0.593880 | 0.539576 |
| 12 | 0.006131 | 1.490619 | 0.564569 | 0.682567 | 0.627583 | 0.564569 | 0.682567 | 0.627583 |
| 13 | 0.001487 | 1.671536 | 0.606713 | 0.733463 | 0.673970 | 0.606713 | 0.733463 | 0.673970 |
| 14 | 0.023817 | 1.810581 | 0.706981 | 0.830453 | 0.774077 | 0.706981 | 0.830453 | 0.774077 |
| 15 | 0.017495 | 1.979867 | 0.750483 | 0.874788 | 0.815630 | 0.750483 | 0.874788 | 0.815630 |

WM: weighted mean; WMD: weighted median.

Table 6. Results of biases given different sample sizes for sample 4

| Sample size $n$ | WM | WMD | $WHL1$ | | | $WHL2$ | | |
|---|---|---|---|---|---|---|---|---|
| | | | $i<j$ | $i\leq j$ | $i<j$ | $i\leq j$ | $i<j$ | $i\leq j$ |
| 3 | 0.003341 | 0.571137 | 0.054455 | 0.190326 | 0.054455 | 0.160538 | 0.270841 | 0.243188 |
| 4 | 0.006683 | 0.735182 | 0.205916 | 0.406355 | 0.406355 | 0.279373 | 0.354942 | 0.300079 |
| 5 | 0.011738 | 0.860739 | 0.445836 | 0.468608 | 0.468608 | 0.374829 | 0.433715 | 0.381896 |
| 6 | 0.005184 | 0.975721 | 0.527651 | 0.631118 | 0.579384 | 0.451484 | 0.504122 | 0.463677 |
| 7 | 0.004441 | 1.065208 | 0.668787 | 0.906668 | 0.820475 | 0.519239 | 0.568104 | 0.535909 |
| 8 | 0.005840 | 1.139922 | 1.004772 | 1.098953 | 1.060279 | 0.581483 | 0.622855 | 0.598111 |
| 9 | 0.013435 | 1.214767 | 1.178325 | 1.286067 | 1.232528 | 0.644959 | 0.680271 | 0.653535 |
| 10 | 0.005792 | 1.291698 | 1.346788 | 1.497167 | 1.411557 | 0.721313 | 0.751430 | 0.725734 |
| 11 | 0.004787 | 1.339866 | 1.589778 | 1.714265 | 1.654797 | 0.762745 | 0.791099 | 0.768321 |
| 12 | 0.012992 | 1.415307 | 1.790764 | 1.908060 | 1.852914 | 0.829621 | 0.855549 | 0.837180 |
| 13 | 0.006141 | 1.460031 | 2.029887 | 2.151688 | 2.091706 | 0.863025 | 0.886453 | 0.870907 |
| 14 | 0.003357 | 1.498904 | 2.253414 | 2.390950 | 2.332289 | 0.913681 | 0.933902 | 0.918962 |
| 15 | 0.007536 | 1.534874 | 2.483312 | 2.607654 | 2.548772 | 0.947184 | 0.965220 | 0.949900 |

WM: weighted mean; WMD: weighted median.

Table 7. Results of biases given different sample sizes for sample 5

| Sample size $n$ | WM | WMD | $WHL1$ | | | $WHL2$ | | |
|---|---|---|---|---|---|---|---|---|
| | | | $i<j$ | $i\leq j$ | $i<j$ | $i\leq j$ | $i<j$ | $i\leq j$ |
| 3 | 0.002277 | 0.013838 | 0.016873 | 0.062361 | 0.016873 | 0.077455 | 0.200847 | 0.138941 |
| 4 | 0.008143 | 0.145553 | 0.060557 | 0.121200 | 0.121200 | 0.215706 | 0.318334 | 0.282437 |
| 5 | 0.001509 | 0.333143 | 0.112806 | 0.116623 | 0.116623 | 0.325929 | 0.414495 | 0.404251 |
| 6 | 0.015216 | 0.516374 | 0.117792 | 0.168326 | 0.143059 | 0.411831 | 0.514329 | 0.503707 |
| 7 | 0.001870 | 0.676427 | 0.154714 | 0.241869 | 0.203409 | 0.543847 | 0.635635 | 0.603226 |
| 8 | 0.002718 | 0.843777 | 0.260223 | 0.302141 | 0.285979 | 0.655762 | 0.754862 | 0.723264 |
| 9 | 0.002947 | 0.990666 | 0.321657 | 0.369027 | 0.343580 | 0.766347 | 0.856898 | 0.825960 |
| 10 | 0.014154 | 1.155812 | 0.395755 | 0.454285 | 0.423593 | 0.882856 | 0.971155 | 0.939988 |
| 11 | 0.013460 | 1.335231 | 0.444625 | 0.499656 | 0.473275 | 1.017763 | 1.104861 | 1.072707 |
| 12 | 0.016229 | 1.496058 | 0.525393 | 0.579390 | 0.553310 | 1.139685 | 1.226501 | 1.194786 |
| 13 | 0.002115 | 1.666903 | 0.622080 | 0.677650 | 0.651009 | 1.250395 | 1.335564 | 1.303952 |
| 14 | 0.009193 | 1.840874 | 0.701369 | 0.758469 | 0.731040 | 1.381030 | 1.465876 | 1.433248 |
| 15 | 0.020972 | 1.975095 | 0.843309 | 0.901667 | 0.874581 | 1.472533 | 1.559089 | 1.525048 |

WM: weighted mean; WMD: weighted median.

Table 8. Results of biases given different sample sizes for sample 6

| Sample size $n$ | WM | WMD | $WHL1$ | | | $WHL2$ | | |
|---|---|---|---|---|---|---|---|---|
| | | | $i < j$ | $i \le j$ | $i < j$ | $i \le j$ | $i < j$ | $i \le j$ |
| 3 | 0.002136 | 0.406575 | 0.001121 | 0.003447 | 0.001121 | 0.039374 | 0.057476 | 0.055461 |
| 4 | 0.010714 | 0.475992 | 0.012424 | 0.017067 | 0.017067 | 0.056247 | 0.046069 | 0.022768 |
| 5 | 0.011286 | 0.505772 | 0.003995 | 0.006285 | 0.006285 | 0.048698 | 0.040305 | 0.026253 |
| 6 | 0.004993 | 0.517687 | 0.003487 | 0.004230 | 0.003859 | 0.038552 | 0.033232 | 0.026058 |
| 7 | 0.001503 | 0.501205 | 0.006287 | 0.005895 | 0.005688 | 0.024021 | 0.020558 | 0.012142 |
| 8 | 0.005429 | 0.488752 | 0.004337 | 0.003847 | 0.004508 | 0.011991 | 0.008261 | 0.000305 |
| 9 | 0.005146 | 0.484854 | 0.001464 | 0.002475 | 0.003025 | 0.013099 | 0.009322 | 0.002399 |
| 10 | 0.002231 | 0.472427 | 0.011861 | 0.013229 | 0.013195 | 0.002280 | 0.001231 | 0.006511 |
| 11 | 0.003350 | 0.483520 | 0.011782 | 0.013313 | 0.014422 | 0.014805 | 0.012872 | 0.007678 |
| 12 | 0.002844 | 0.470293 | 0.006605 | 0.006250 | 0.006258 | 0.009589 | 0.007891 | 0.003086 |
| 13 | 0.008720 | 0.435471 | 0.006952 | 0.008108 | 0.007078 | 0.003900 | 0.005804 | 0.009835 |
| 14 | 0.004399 | 0.454714 | 0.001435 | 0.003075 | 0.002824 | 0.015838 | 0.014597 | 0.010588 |
| 15 | 0.013154 | 0.424622 | 0.008014 | 0.005850 | 0.006492 | 0.003436 | 0.004679 | 0.008141 |

WM: weighted mean; WMD: weighted median.

After that, the relative efficiencies are also calculated and provided in Tables 9-13 after utilizing the methods of the weighted mean, the weighted median, and the proposed WHL (i.e., $WHL1$ and $WHL2$) estimator. As shown in Tables 9-13, the proposed $WHL2$ estimators always keep a high relative efficiency. Specifically, the proposed $WHL1$ and $WHL2$ estimators commonly outperform the weighted median in Tables 9 and 10 because they have a similar relative efficiency. In Tables 11-13, the relative efficiencies of $WHL2$ estimators are decreasing as the sample size grows, but they are still and commonly greater than the weighted median and $WHL1$ estimators. It can be concluded that $WHL2$ estimators need to be preferred to estimate the location parameter.

Table 9. Results of relative efficiencies given different sample sizes for sample 2

| Sample size $n$ | WM | WMD | $WHL1$ | | | $WHL2$ | | |
|---|---|---|---|---|---|---|---|---|
| | | | $i < j$ | $i \le j$ | $i < j$ | $i \le j$ | $i < j$ | $i \le j$ |
| 3 | 100.22463 | 73.81635 | 92.50902 | 97.88818 | 92.50902 | 92.50902 | 97.88818 | 92.50902 |
| 4 | 98.08601 | 82.38248 | 98.08601 | 89.55031 | 89.55031 | 98.08601 | 89.55031 | 89.55031 |
| 5 | 99.4501 | 69.38295 | 93.70788 | 92.60952 | 92.60952 | 93.70788 | 92.60952 | 92.60952 |
| 6 | 100.66316 | 77.95909 | 94.28057 | 92.8731 | 94.35512 | 94.28057 | 92.8731 | 94.35512 |
| 7 | 100.31067 | 68.91528 | 94.73616 | 92.7345 | 93.29221 | 94.73616 | 92.7345 | 93.29221 |
| 8 | 102.7107 | 75.83573 | 96.97018 | 95.92923 | 96.15034 | 96.97018 | 95.92923 | 96.15034 |
| 9 | 99.62647 | 65.67844 | 94.2485 | 93.04202 | 93.70977 | 94.2485 | 93.04202 | 93.70977 |
| 10 | 101.27969 | 72.49826 | 95.21049 | 93.81953 | 94.81572 | 95.21049 | 93.81953 | 94.81572 |
| 11 | 97.63992 | 64.88339 | 91.91425 | 91.0391 | 91.31824 | 91.91425 | 91.0391 | 91.31824 |
| 12 | 99.86891 | 70.27851 | 95.04642 | 93.74014 | 94.46938 | 95.04642 | 93.74014 | 94.46938 |
| 13 | 99.21092 | 66.01404 | 92.89359 | 92.04132 | 92.37187 | 92.89359 | 92.04132 | 92.37187 |
| 14 | 101.94005 | 70.86857 | 96.53721 | 95.55961 | 96.12655 | 96.53721 | 95.55961 | 96.12655 |
| 15 | 100.42115 | 64.39188 | 95.42314 | 94.28905 | 94.82625 | 95.42314 | 94.28905 | 94.82625 |

WM: weighted mean; WMD: weighted median.

Table 10. Results of relative efficiencies given different sample sizes for sample 3

| Sample size $n$ | WM | WMD | $WHL1$ | | | $WHL2$ | | |
|---|---|---|---|---|---|---|---|---|
| | | | $i<j$ | $i\leq j$ | $i<j$ | $i\leq j$ | $i<j$ | $i\leq j$ |
| 3 | 100.14774 | 87.67921 | 79.21749 | 105.2279 | 79.21749 | 79.21749 | 105.2279 | 79.21749 |
| 4 | 98.68856 | 100.4415 | 98.68856 | 101.5152 | 101.5152 | 98.68856 | 101.5152 | 101.5152 |
| 5 | 99.96807 | 82.20132 | 100.5852 | 100.6144 | 100.6144 | 100.5852 | 100.6144 | 100.6144 |
| 6 | 99.64699 | 90.6188 | 95.15476 | 98.38767 | 97.77224 | 95.15476 | 98.38767 | 97.77224 |
| 7 | 97.47182 | 74.12777 | 92.41489 | 96.87102 | 95.11913 | 92.41489 | 96.87102 | 95.11913 |
| 8 | 99.58906 | 80.85296 | 101.3963 | 101.2256 | 101.4464 | 101.3963 | 101.2256 | 101.4464 |
| 9 | 99.85738 | 69.24401 | 98.95121 | 98.0606 | 98.13669 | 98.95121 | 98.0606 | 98.13669 |
| 10 | 99.82555 | 72.98319 | 97.83832 | 97.81601 | 98.0697 | 97.83832 | 97.81601 | 98.0697 |
| 11 | 99.4304 | 62.9157 | 96.40107 | 96.02543 | 96.37101 | 96.40107 | 96.02543 | 96.37101 |
| 12 | 96.92674 | 64.64522 | 94.12186 | 92.8688 | 93.64005 | 94.12186 | 92.8688 | 93.64005 |
| 13 | 99.36175 | 58.83056 | 96.41901 | 94.60463 | 95.60709 | 96.41901 | 94.60463 | 95.60709 |
| 14 | 98.18103 | 59.70523 | 93.80456 | 92.08762 | 93.13938 | 93.80456 | 92.08762 | 93.13938 |
| 15 | 98.05116 | 53.74487 | 92.85806 | 90.87135 | 91.89797 | 92.85806 | 90.87135 | 91.89797 |

WM: weighted mean; WMD: weighted median.

Table 11. Results of relative efficiencies given different sample sizes for sample 4

| Sample size $n$ | WM | WMD | $WHL1$ | | | $WHL2$ | | |
|---|---|---|---|---|---|---|---|---|
| | | | $i<j$ | $i\leq j$ | $i<j$ | $i\leq j$ | $i<j$ | $i\leq j$ |
| 3 | 100.0416 | 61.16901 | 94.86973 | 77.07104 | 94.86973 | 92.30774 | 81.28131 | 84.90442 |
| 4 | 99.51343 | 48.7192 | 79.88857 | 49.01845 | 49.01845 | 81.00594 | 72.25766 | 76.28216 |
| 5 | 99.17076 | 40.9165 | 47.79451 | 45.29223 | 45.29223 | 71.50564 | 66.02359 | 70.64913 |
| 6 | 103.82044 | 35.70755 | 42.93895 | 37.98992 | 40.80616 | 66.53338 | 61.65766 | 65.34185 |
| 7 | 98.89074 | 31.33203 | 35.92066 | 26.39867 | 28.88051 | 60.03986 | 56.14569 | 58.41064 |
| 8 | 101.87601 | 28.23734 | 24.5483 | 22.0484 | 22.98559 | 55.15699 | 52.11132 | 53.58221 |
| 9 | 100.59514 | 25.88307 | 20.29142 | 18.12807 | 19.11399 | 51.0733 | 48.7758 | 50.42444 |
| 10 | 99.31157 | 23.38479 | 17.16144 | 14.73269 | 16.17728 | 46.1135 | 44.43866 | 45.94301 |
| 11 | 100.42417 | 22.00475 | 13.70248 | 12.29117 | 12.92062 | 43.74432 | 42.22379 | 43.48983 |
| 12 | 102.64261 | 20.58355 | 11.59354 | 10.53095 | 11.00469 | 40.97446 | 39.65186 | 40.57443 |
| 13 | 98.705567 | 19.10543 | 9.638392 | 8.78269 | 9.19125 | 38.55901 | 37.45140 | 38.13773 |
| 14 | 100.74523 | 18.27410 | 8.28887 | 7.489572 | 7.809517 | 36.27889 | 35.39965 | 36.04007 |
| 15 | 99.772476 | 17.72554 | 7.094203 | 6.536302 | 6.792065 | 35.04917 | 34.33173 | 34.97389 |

WM: weighted mean; WMD: weighted median.

Table 12. Results of relative efficiencies given different sample sizes for sample 5

| Sample size $n$ | WM | WMD | $WHL1$ | | | $WHL2$ | | |
|---|---|---|---|---|---|---|---|---|
| | | | $i<j$ | $i\le j$ | $i<j$ | $i\le j$ | $i<j$ | $i\le j$ |
| 3 | 101.71092 | 99.15974 | 94.58614 | 86.62312 | 94.58614 | 105.6430 | 112.0046 | 111.1938 |
| 4 | 100.97126 | 93.66929 | 92.58549 | 71.34274 | 71.34274 | 114.0131 | 108.1108 | 108.7297 |
| 5 | 101.50531 | 86.56135 | 75.54125 | 73.44196 | 73.44196 | 108.5713 | 104.2459 | 104.1419 |
| 6 | 100.80025 | 74.40267 | 74.59482 | 67.62586 | 71.92957 | 105.1041 | 93.74823 | 97.60468 |
| 7 | 99.99789 | 67.32531 | 69.18725 | 59.91788 | 63.37215 | 93.09562 | 84.76996 | 86.82905 |
| 8 | 99.65877 | 59.29355 | 61.07967 | 57.19438 | 58.32646 | 84.56199 | 76.53326 | 79.08046 |
| 9 | 101.04617 | 53.27858 | 57.86424 | 53.30943 | 55.5614 | 78.21559 | 70.93321 | 73.22003 |
| 10 | 99.11911 | 46.06184 | 54.13815 | 49.48179 | 52.21207 | 70.05159 | 63.92879 | 66.00857 |
| 11 | 102.50873 | 40.53399 | 51.37758 | 48.02073 | 49.55277 | 62.9932 | 57.56659 | 59.43875 |
| 12 | 100.78422 | 35.5245 | 47.92536 | 44.91139 | 46.33103 | 56.00456 | 51.36637 | 52.97324 |
| 13 | 101.55815 | 31.62712 | 45.28078 | 42.1864 | 43.62824 | 51.39535 | 47.32896 | 48.77728 |
| 14 | 100.74512 | 28.30968 | 41.74204 | 39.15376 | 40.41341 | 46.39977 | 42.8751 | 44.15598 |
| 15 | 101.77869 | 26.06913 | 38.39224 | 35.99221 | 37.08667 | 43.29751 | 40.081 | 41.3092 |

WM: weighted mean; WMD: weighted median.

Table 13. Results of relative efficiencies given different sample sizes for sample 6

| Sample size $n$ | WM | WMD | $WHL1$ | | | $WHL2$ | | |
|---|---|---|---|---|---|---|---|---|
| | | | $i<j$ | $i\le j$ | $i<j$ | $i\le j$ | $i<j$ | $i\le j$ |
| 3 | 100.09982 | 71.21666 | 96.33645 | 84.14587 | 96.33645 | 95.81054 | 87.44182 | 90.38575 |
| 4 | 99.73781 | 60.84846 | 88.81656 | 68.40871 | 68.40871 | 91.07134 | 83.38012 | 85.41539 |
| 5 | 101.67489 | 55.06522 | 72.22204 | 69.69053 | 69.69053 | 86.26954 | 79.99928 | 83.33422 |
| 6 | 100.5992 | 51.6187 | 71.78459 | 65.96182 | 69.56436 | 82.49755 | 77.73682 | 80.06176 |
| 7 | 99.72804 | 50.6267 | 67.45218 | 60.10107 | 62.64779 | 79.01362 | 74.97096 | 76.55964 |
| 8 | 98.49996 | 51.02162 | 61.85429 | 58.63361 | 59.69813 | 78.1008 | 74.94932 | 76.16586 |
| 9 | 103.1819 | 50.49562 | 62.31883 | 58.55635 | 60.57065 | 78.08748 | 75.22478 | 76.46025 |
| 10 | 100.28744 | 51.31107 | 59.69346 | 56.262 | 58.32388 | 75.71946 | 73.32832 | 74.40592 |
| 11 | 99.65035 | 48.88387 | 55.11792 | 52.38316 | 53.80658 | 72.75247 | 70.72123 | 71.6444 |
| 12 | 100.04079 | 48.86973 | 53.74931 | 51.00423 | 52.25708 | 71.05954 | 69.36237 | 70.05226 |
| 13 | 101.37332 | 52.3623 | 51.3656 | 48.71281 | 49.90079 | 73.68474 | 72.14918 | 72.79458 |
| 14 | 99.57687 | 50.27895 | 50.35117 | 48.17385 | 49.19634 | 70.19349 | 68.86469 | 69.47928 |
| 15 | 98.66607 | 51.41945 | 46.96079 | 44.90419 | 45.81284 | 69.13137 | 67.99888 | 68.53007 |

WM: weighted mean; WMD: weighted median.

### 4.3 Sensitivity study

In addition to the above 6 small-size samples, this study tests 12 larger-size datasets (i.e., Cases 1-12, see Table 14) generated using different distributions (i.e., uniform, normal, chi-square, and Poisson) with weights to fairly evaluate their sensitivity to different outlier proportions. Notably, three different weight constructions are considered, where the first one is random and unordered, and the second and third are random and ordered, in descending and ascending orders, respectively.

Table14 Observation and weight constructions under consideration

| Case ID | Observation constructions | | | Weight constructions | Outlier proportion range |
|---|---|---|---|---|---|
| | Distribution type | Number of observations | Parameters | | |
| 1 | Uniform | 100 | $U_1\sim U$ (50, 150) | W1: $U\sim U$ (10, 100) | 0~25% |
| 2 | | 100 | $U_2\sim$ U (50, 150) | W2: $U_2/100$ | 0~25% |
| 3 | | 100 | $U_3\sim$ U (50, 150) | W3: 3-$U_3/100$ | 0~25% |
| 4 | Normal | 100 | $N_1\sim$ N (100, 20) | W1: $U\sim$ U (10, 100) | 0~25% |
| 5 | | 100 | $N_2\sim$ N (100, 20) | W2: $N_2/100$ | 0~25% |
| 6 | | 100 | $N_3\sim$ N (100, 20) | W3: 3-$N_3/100$ | 0~25% |
| 7 | Chi-square | 100 | $C_1\sim\chi^2(100, 1)$ | W1: $U\sim$ U (10, 100) | 0~25% |
| 8 | | 100 | $C_2\sim\chi^2(100, 1)$ | W2: $C_2/100$ | 0~25% |
| 9 | | 100 | $C_3\sim\chi^2(100, 1)$ | W3: 3-$C_3/100$ | 0~25% |
| 10 | Poisson | 100 | $P_1\sim P(100)$ | W1: $U\sim$ U (10, 100) | 0~25% |
| 11 | | 100 | $P_2\sim P(100)$ | W2: $P_2/100$ | 0~25% |
| 12 | | 100 | $P_3\sim P(100)$ | W3: 3-$P_3/100$ | 0~25% |

Using the above observations and weight constructions, this study first derives the bias of the proposed estimators without considering any outliers. As presented in Figs. 3–5, the Bias of different estimators (weighted mean, $HL$, $WHL1$, and $WHL2$) were visually compared and illustrated under three weight constructions (W1, W2, and W3 in Table 14). From Fig. 3, it is observed that all estimators are close to zero when the weights are random and unordered. However, when the weights are ordered, obvious visual differences are observed, as shown in Figs. 4 and 5. Specifically, the larger the observation value is, the greater the weight is. As shown in Fig. 4, the Bias of the proposed $WHL1$ and $WHL2$ estimators are smaller than those of the existing estimator ($HL$). After comparing their Biases, it is found that the $WHL2$ estimators outperform the $WHL1$ estimators. A similar tendency can be identified in Fig. 5, wherein the weights are inversely related to the observation values. This is strong evidence that the proposed robust estimators ($WHL1$ and $WHL2$) are much closer to the weighted mean and more reliable for substituting the weighted mean when no outliers are involved.

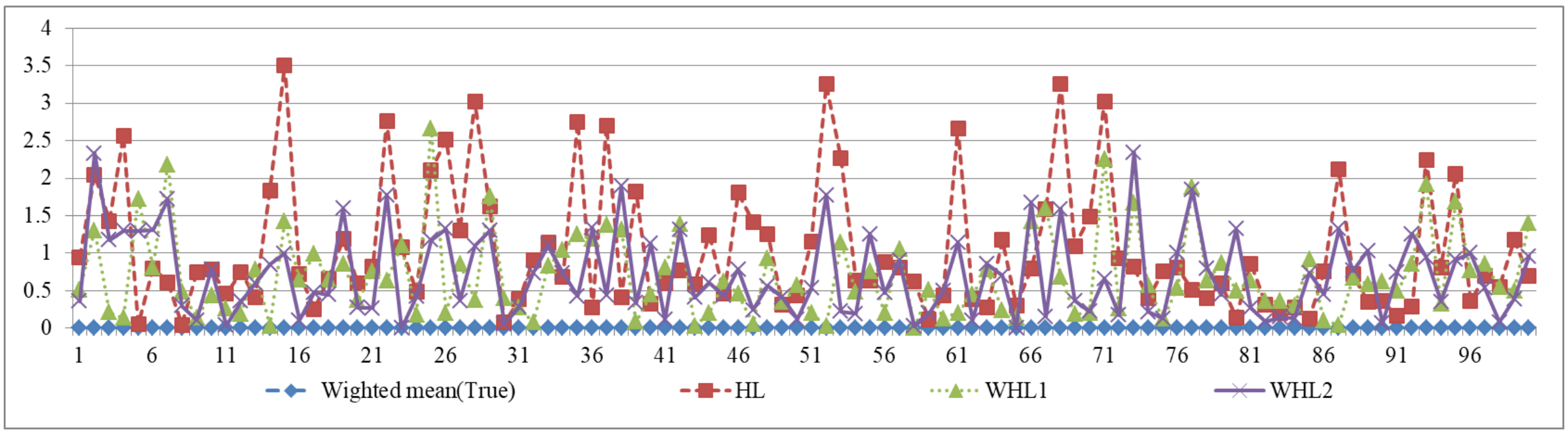

(a) $U_1$~ U (50, 150), W1: U~ U (10, 100)

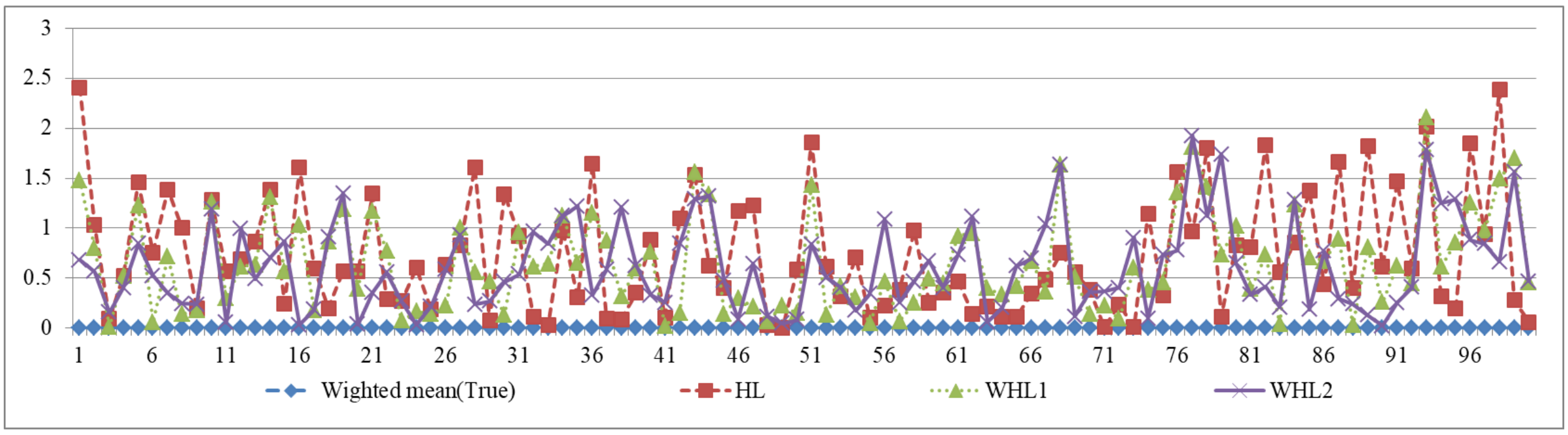

(b) $N_1$~ N (100, 20), W1: U~ U (10, 100)

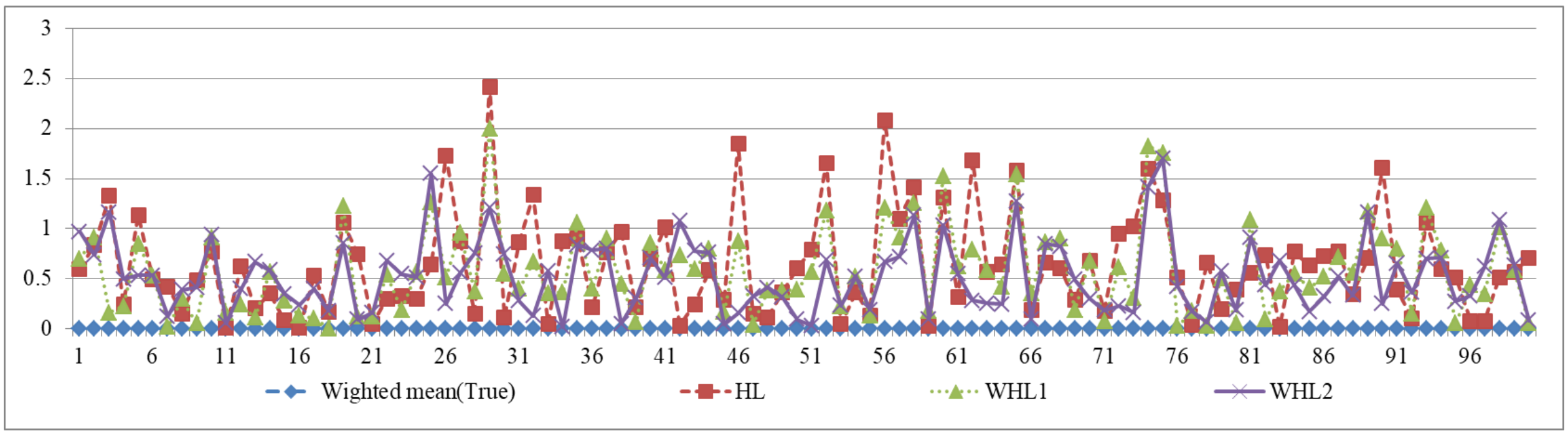

(c) $C_1 \sim \chi^2(100,1)$, W1: U~ U (10, 100)

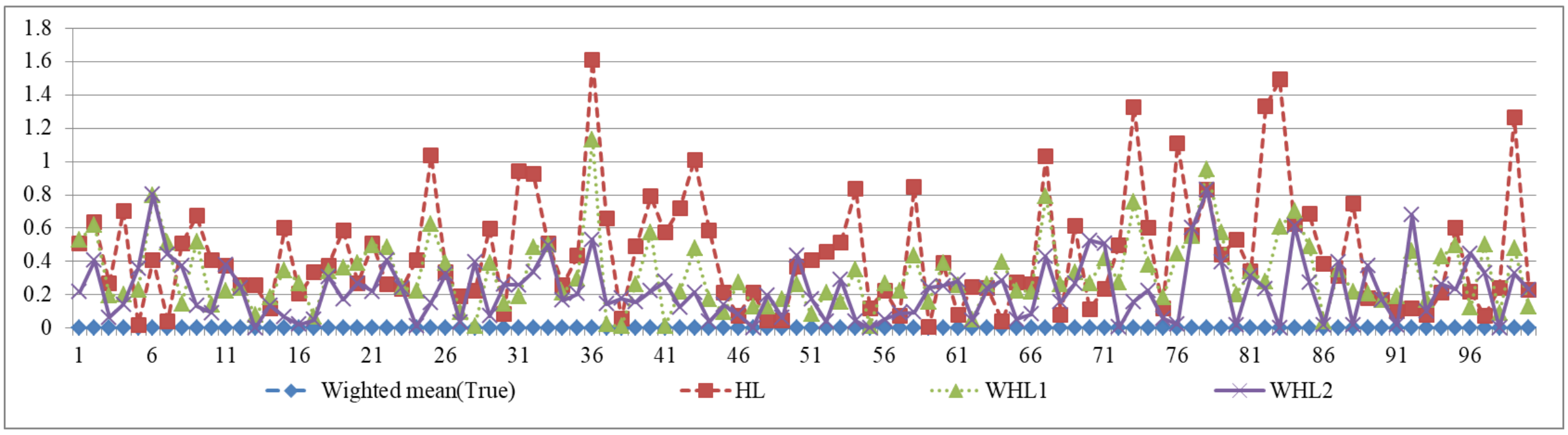

(d) $P_1 \sim P(100)$, W1: U~ U (10, 100)

Fig. 3. Comparison of Bias using different estimators under random and unordered weights

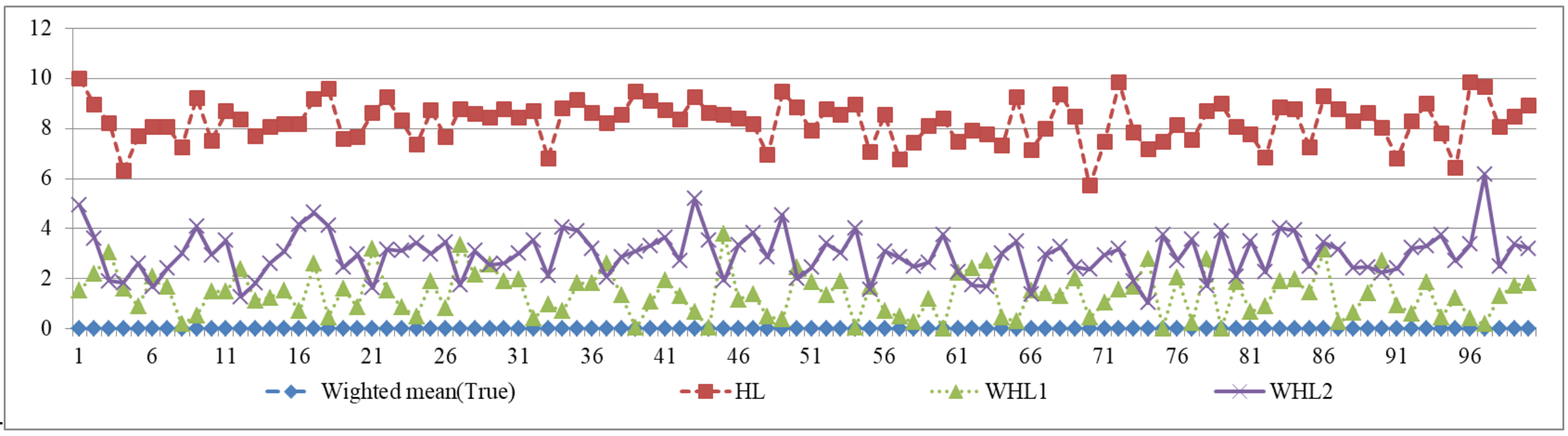

4

(a) $U_2 \sim$ U (50, 150), W2: $U_2/100$

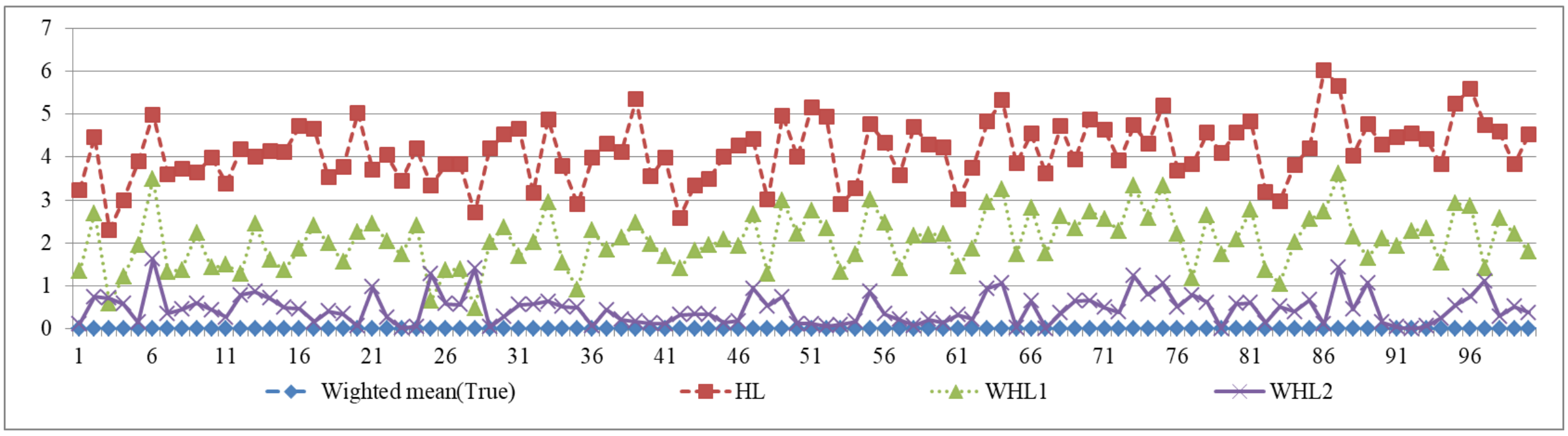

(b) $N_2 \sim$ N (100, 20), W2: $N_2/100$

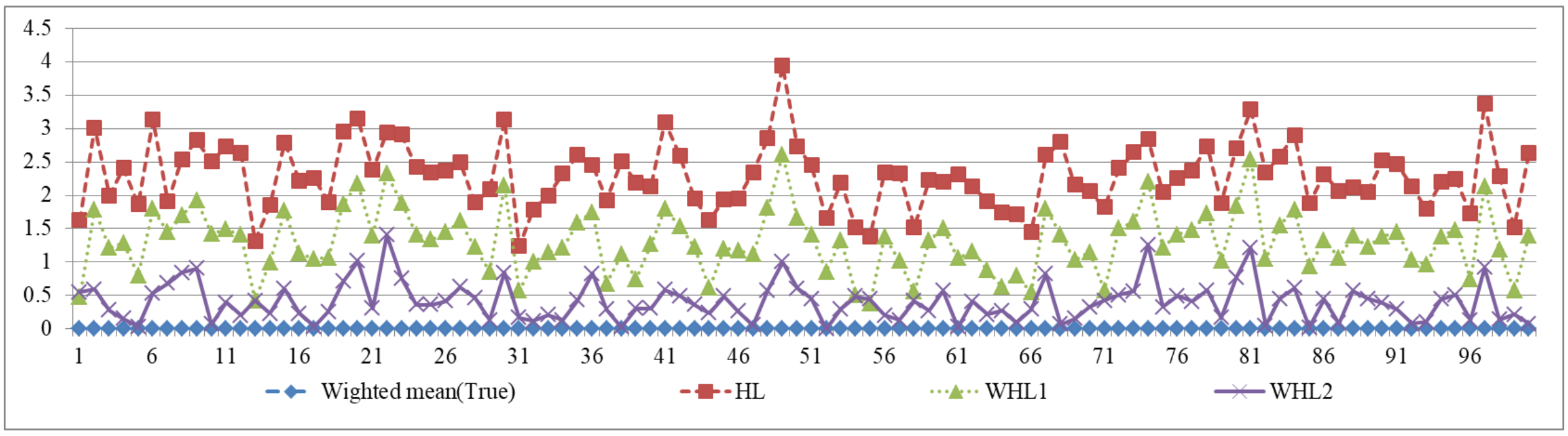

(c) $C_2 \sim \chi^2(100, 1)$, W2: $C_2/100$

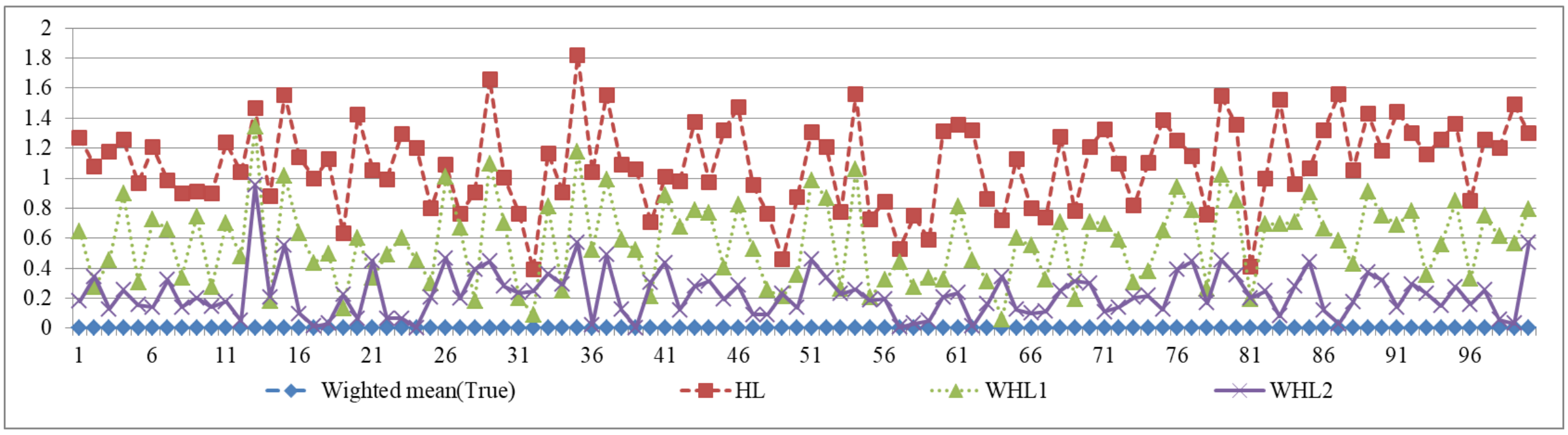

(d) $P_2 \sim P(100)$, W2: $P_2/100$

Fig. 4. Comparison of Bias using different estimators under random and ascending-order weights

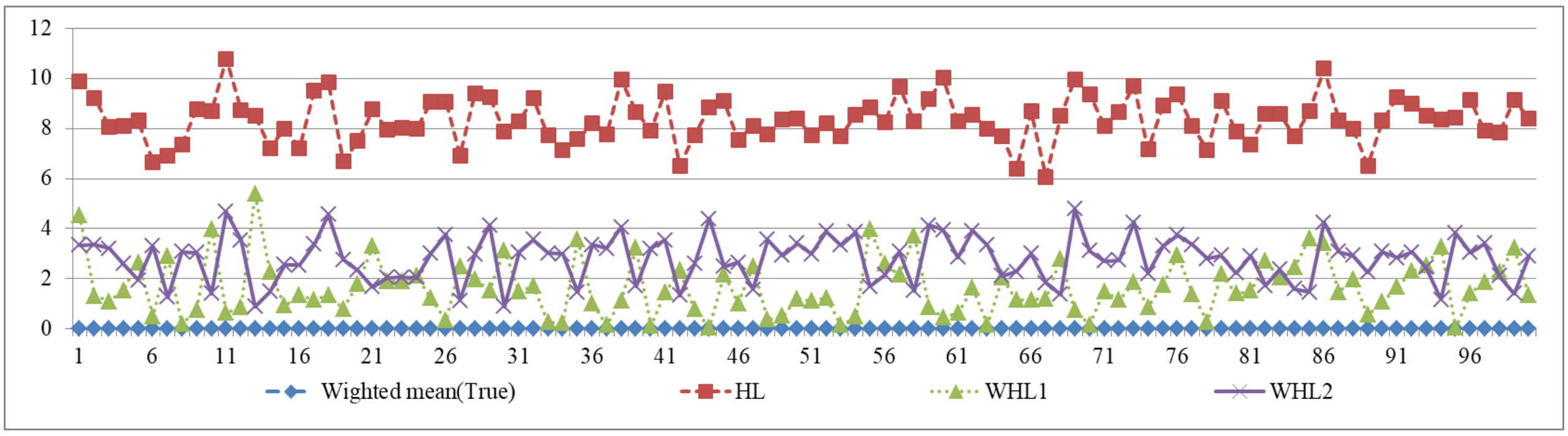

(a) $U_3 \sim$ U (50, 150), W3: 3-$U_3/100$

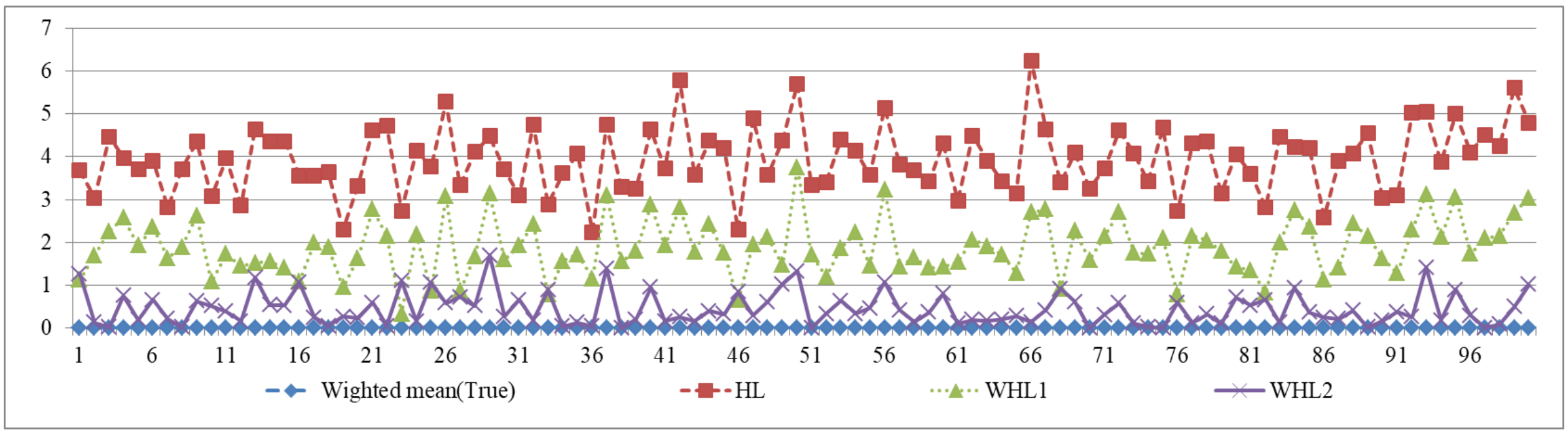

(b) $N_3 \sim$ N (100, 20), W3: 3-$N_3/100$

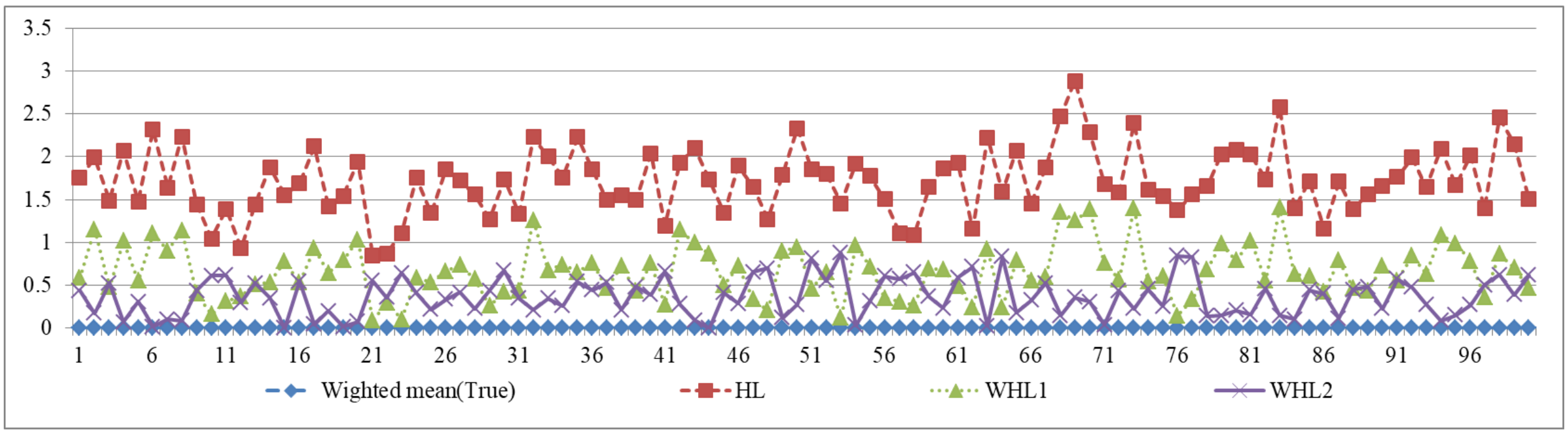

(c) $C_3 \sim \chi^2(100, 1)$, W3: 3-$C_3/100$

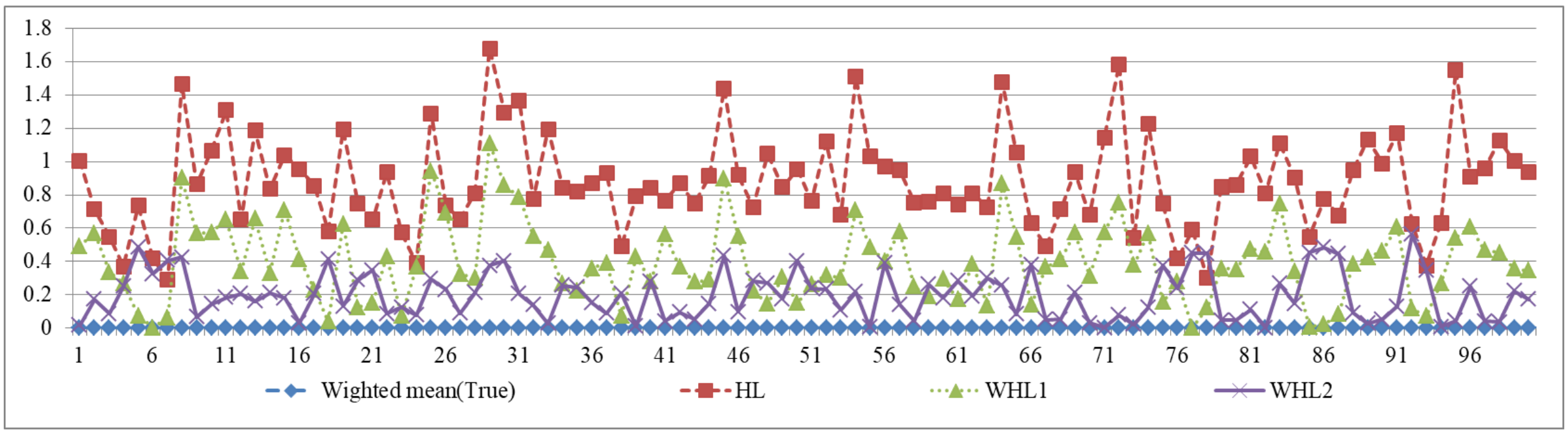

(d) $P_3 \sim P(100)$, W3: 3-$P_3/100$

Fig. 5. Comparison of Bias using different estimators under random and descending-order weights

Next, this study investigates the robustness of the proposed estimators, given different outlier proportions. As illustrated in Figs. 6–9, when outliers are involved, the outlier-resistant behaviors of the weighted mean, $HL$, $WHL1$, and $WHL2$ are visually compared. Clearly, the average bias using the weighted mean is zero without considering any outliers. However, when outliers are considered, the weighted mean shifts away from zero. All the average biases using robust estimators remain smaller than the weighted mean when the outlier proportion increases from 0 to 25%, indicating that the robust estimators applied in this study are much more reliable than the weighted mean when the outliers are involved. In particular, when the weights are random and unordered, it is difficult to determine which model has better robust performance because the weights tend to be more symmetric [see Figs. 6-9 (a)]. However, in practice, the weights are more likely to be nonsymmetric. Thus, the newly proposed robust estimators ($WHL1$ and $WHL2$) outperform $HL$ when the weights are ordered [see Figs. 6-9 (b) and (c)]. The proportion of outliers is typically very small. The above discussion provides strong evidence to verify the theoretical contributions of the robust optimization models proposed in this study.

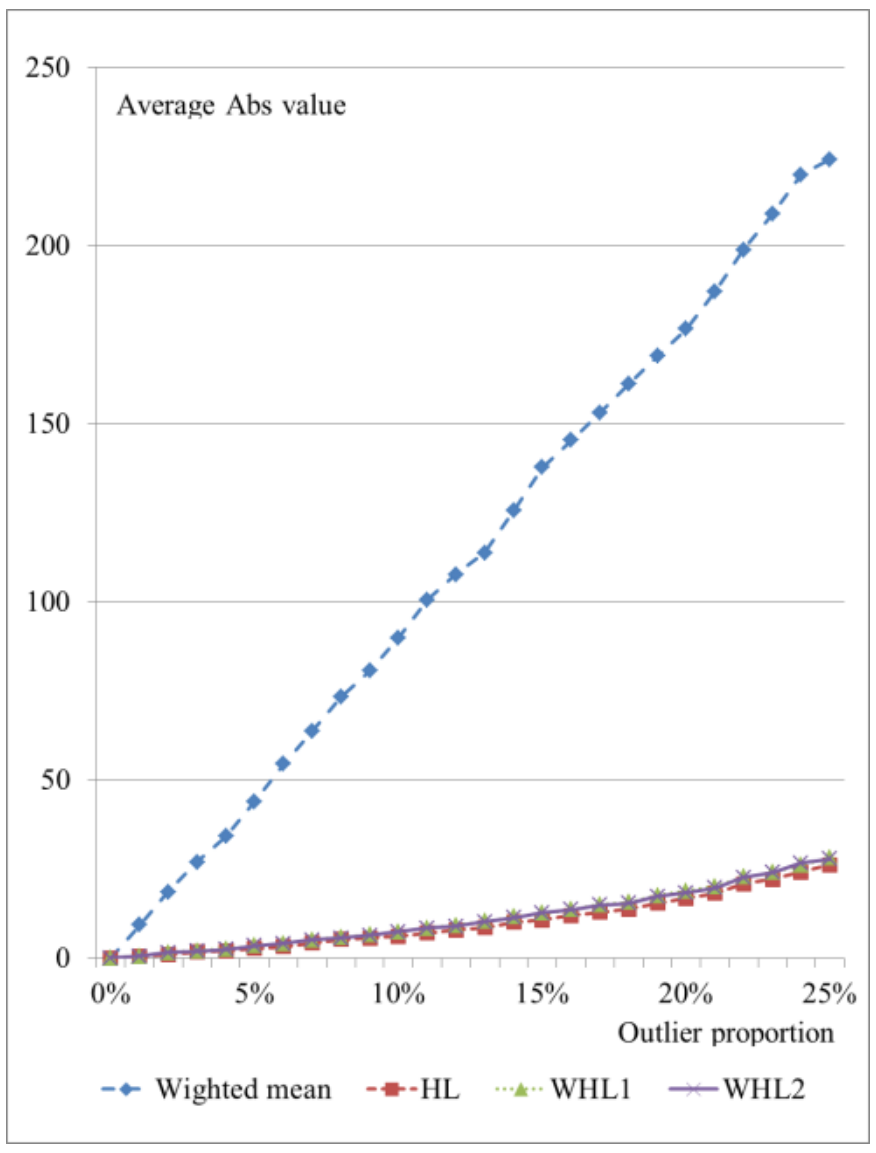

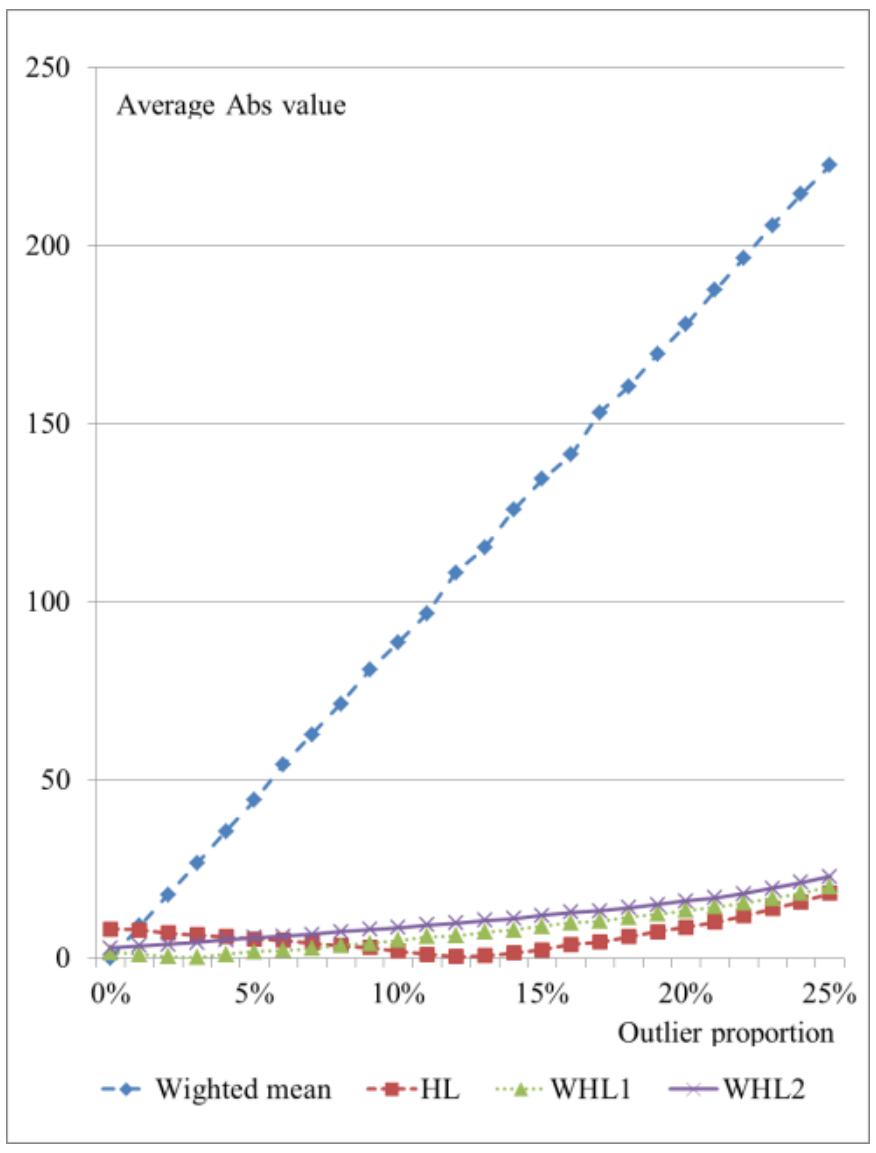

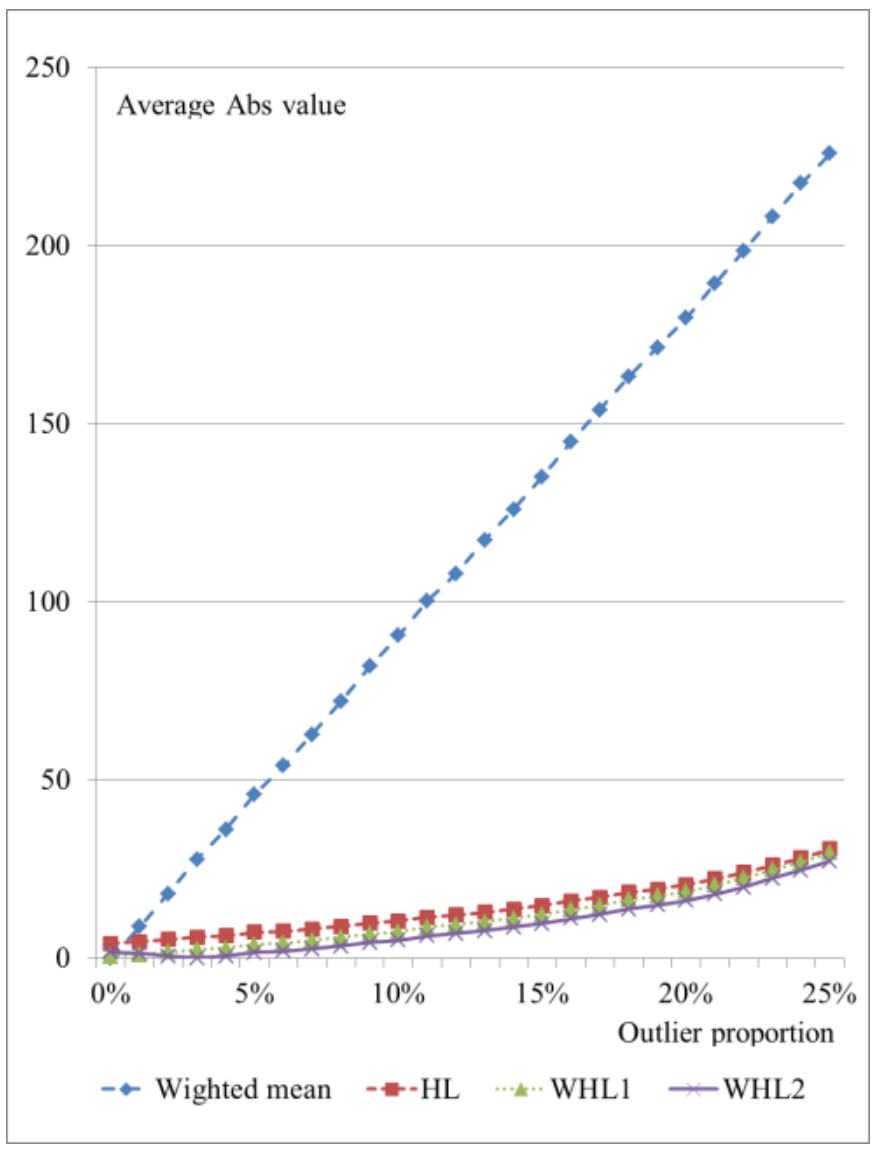

(a) W1 (b) W2 (c) W3

Fig. 6. Comparison of average bias using different estimators with uniform distribution

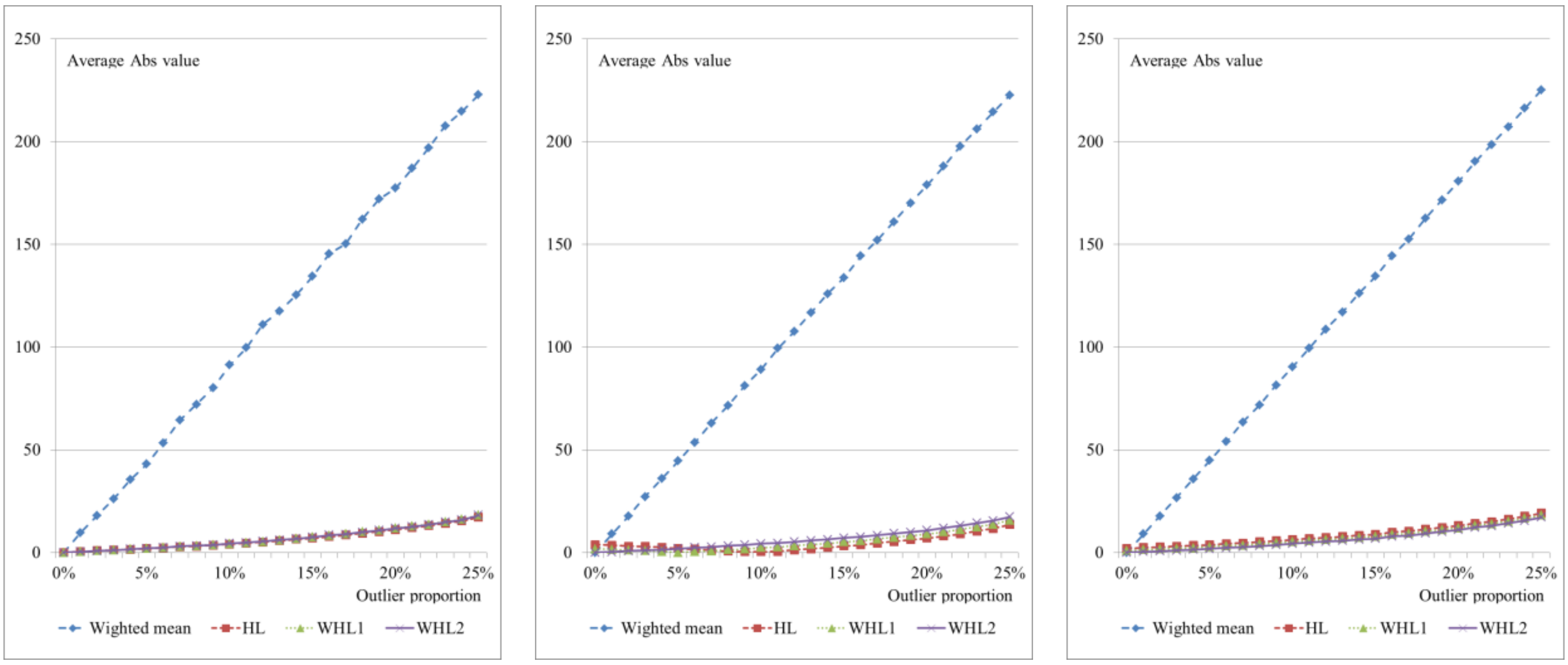

(a) W1 (b) W2 (c) W3

Fig. 7. Comparison of average bias using different estimators with normal distribution

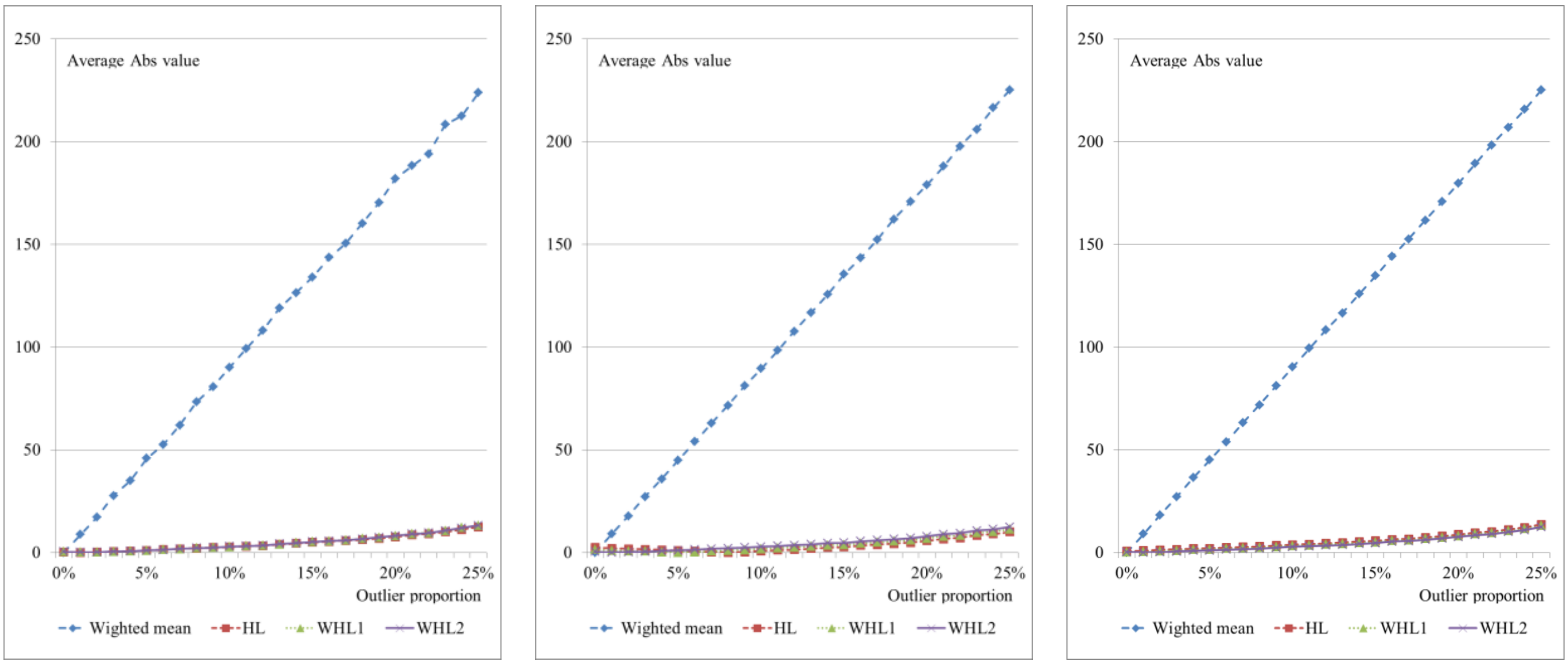

(a) W1 (b) W2 (c) W3

Fig. 8. Comparison of average bias using different estimators with Chi-square distribution

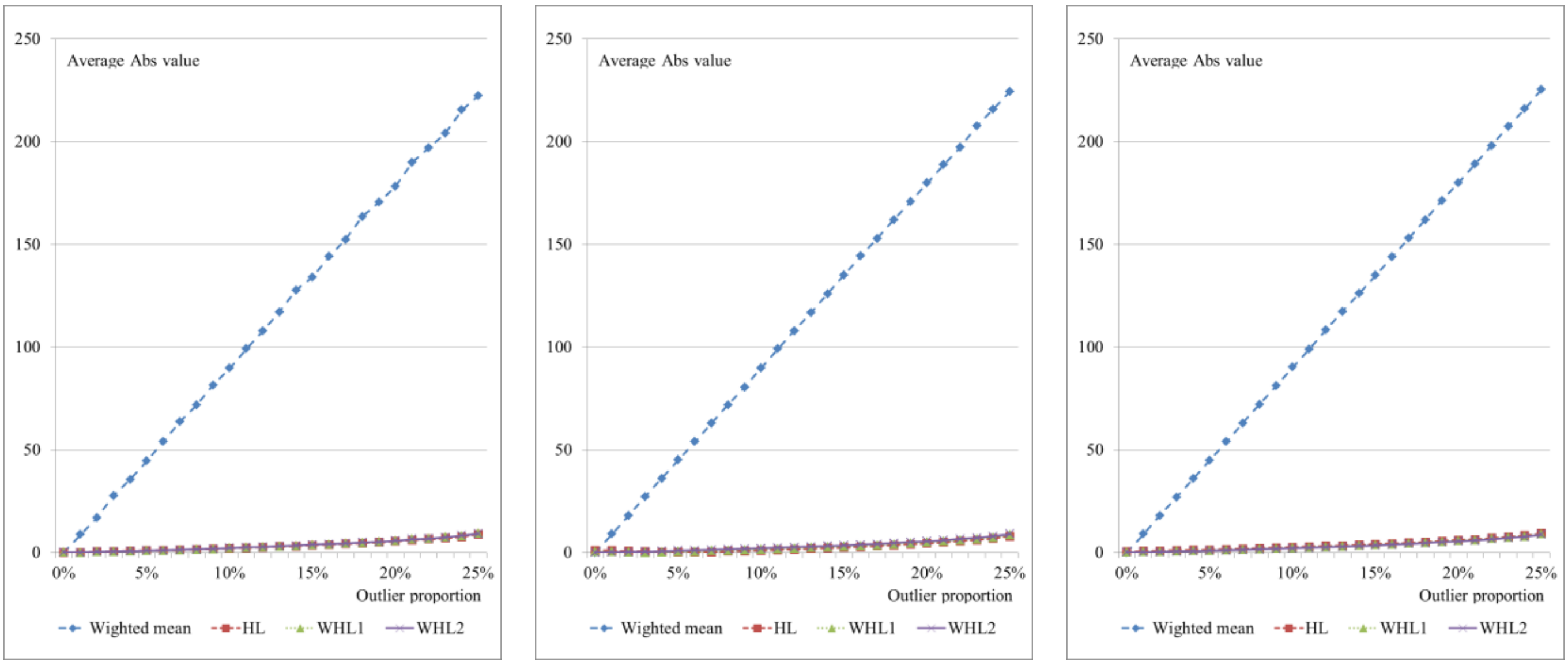

(a) W1 (b) W2 (c) W3

Fig. 9. Comparison of average bias using different estimators with Poisson distribution

# 5 Conclusions

In this paper, we proposed two new categories of WHL estimators for robust location estimation when the sample data contains weights, where the first category of WHL estimators (i.e., $WHL1$) is defined as the median of all pairwise weighted averages and the second WHL estimators (i.e., $WHL2$) is defined as the weighted median of all pairwise weighted averages. Then, this study investigated their robust properties and obtained the exact finite-sample breakdown points of the $WHL1$ estimator and closed-form finite-sample breakdown points of the $WHL2$ estimator. After that, the newly proposed WHL estimator was compared with the traditional ones in terms of bias and relative efficiency through extensive Monte Carlo simulations under different sample sizes, weight configurations, and data distributions. The simulation results reveal that the newly proposed $WHL1$ and $WHL2$ estimators obtain markedly lower bias than the conventional location estimators and the relative efficiency of $WHL2$ estimators is higher than that of the weighted median and $WHL1$ estimators in most cases. Through sensitivity analysis, it is found that the newly proposed $WHL1$ and $WHL2$ estimators are much closer to the weighted mean and more reliable for substituting the weighted mean when no outliers are involved. The newly proposed $WHL1$ and $WHL2$ also remain stable in robustness compared with the $HL$ estimator in the presence of contaminated data.

In addition to the aforementioned contributions, there are two potential directions worth investigating in the future. It would be interesting to combine the proposed robust estimators and conventional ones to develop some more reliable location estimators. Another potential direction is to investigate the exact breakdown points of the newly proposed $WHL2$ estimators.

## Acknowledgments

This work was supported by National Natural Science Foundation of China (No.72104020) and National Research Foundation of Korea (NRF) grant funded by the Korea government (MSIT) (No. 2022R1A2C1091319 and RS-2023-00242528).